\documentclass[pdflatex,sn-mathphys-num,iicol]{sn-jnl}
\usepackage{graphicx}
\usepackage{algorithm}
\usepackage{algorithmicx}
\usepackage[noend]{algpseudocode}
\usepackage{array}
\usepackage{amsfonts}
\usepackage{multirow}
\usepackage{makecell}
\usepackage{bm}
\usepackage{amsthm}
\usepackage{setspace}
\usepackage{marvosym}
\usepackage{xspace}
\usepackage{enumitem}
\usepackage{booktabs}
\usepackage{adjustbox}
\setlist{leftmargin=7mm}
\usepackage{hyperref}

\usepackage{color}
\usepackage{xcolor}
\usepackage{colortbl}
\usepackage{pifont}

\long\def\comment#1{}

\def\ie{$i.e.$}
\def\eg{$e.g.$}

\newcommand{\partitle}[1]{\vspace{0.3em} \noindent \textbf{#1.}}
\newcommand{\firstpartitle}[1]{\vspace{0em} \noindent \textbf{#1.}}
\newcommand{\tightpartitle}[1]{\vspace{0.3em} \noindent \textbf{#1.}}

\newcommand{\sys}{{\sc DATABench}\xspace}
\hypersetup{hidelinks,
	colorlinks=true,
	linkcolor=black,
        filecolor=black,      
        urlcolor=blue,
	pdfstartview=Fit,
        citecolor=blue,
	breaklinks=true
}

\definecolor{success}{HTML}{BBDEFB}
\definecolor{failure}{HTML}{FFAB91}
\newcommand{\cmark}{\textcolor{green!60!black}{\checkmark}}  
\newcommand{\xmark}{\textcolor{red}{\ding{55}}}
\newcommand{\cellgreen}{\cellcolor{success!80}}
\newcommand{\cellred}{\cellcolor{failure!80}}
\newcommand{\successcolor}{\textcolor{success}{blue}\xspace}
\newcommand{\failurecolor}{\textcolor{failure}{orange}\xspace}

\newtheorem{definition}{Definition}

\usepackage{tcolorbox}

\newtcolorbox{takeawaybox}{
  colback=gray!20,
  colframe=gray!20,
  coltitle=black,
  arc=4pt,
  boxrule=0.5pt,
  boxsep=2pt,
  left=2pt,
  right=2pt,
  top=2pt,
  bottom=2pt,
  before skip=0.5\baselineskip,
  after skip=0.5\baselineskip
}

\usepackage[cmex10]{amsmath}
\usepackage{caption}
\usepackage[font=footnotesize]{subfig}

\begin{document}
%
\title{\sys: Evaluating Dataset Auditing in Deep Learning from an Adversarial Perspective}



\author[1]{\fnm{Shuo}\sur{Shao}}\email{shaoshuo\_ss@zju.edu.cn}
\author*[3]{\fnm{Yiming}\sur{Li}}\email{liyiming.tech@gmail.com}
\author[4]{\fnm{Mengren}\sur{Zheng}}\email{mengrenzheng@gmail.com}
\author[1]{\fnm{Zhiyang}\sur{Hu}}\email{zhiyanghu@foxmail.com}
\author[1]{\fnm{Yukun}\sur{Chen}}\email{yukunchen@zju.edu.cn}
\author[3]{\fnm{Boheng}\sur{Li}}\email{randy.bh.li@foxmail.com}
\author[1]{\fnm{Yu}\sur{He}}\email{yuherin@zju.edu.cn}
\author[5]{\fnm{Junfeng}\sur{Guo}}\email{gjf2023@umd.edu}
\author[3]{\fnm{Dacheng}\sur{Tao}}\email{dacheng.tao@ntu.edu.sg}
\author[1,2]{\fnm{Zhan}\sur{Qin}}\email{qinzhan@zju.edu.cn}

\affil[1]{\orgname{State Key Laboratory of Blockchain and Data Security, Zhejiang University}, \orgaddress{ \city{Hangzhou}, \postcode{310027}, \country{China}}}
\affil[2]{\orgname{Hangzhou High-Tech Zone (Binjiang) Institute of Blockchain and Data Security}, \orgaddress{ \city{Hangzhou}, \postcode{310051}, \country{China}}}
\affil[3]{ \orgname{Nanyang Technological University}, \orgaddress{ \city{Singapore}, \postcode{639798}, \country{Singapore}}}
\affil[4]{\orgname{Chongqing University}, \orgaddress{ \city{Chongqing}, \postcode{400044}, \country{China}}}
\affil[5]{\orgname{University of Maryland at College Park}, \orgaddress{ \city{College Park}, \postcode{20742}, \country{USA}}}

\abstract{
The widespread application of Deep Learning across diverse domains hinges critically on the quality and composition of training datasets. However, the common lack of disclosure regarding their usage raises significant privacy and copyright concerns. Dataset auditing, which aims to determine if a specific dataset was used to train a given suspicious model, provides promising solutions to addressing these transparency gaps. While prior work has developed various auditing methods, their resilience against dedicated adversarial attacks remains largely unexplored. To bridge the gap, this paper initiates a comprehensive study evaluating dataset auditing from an adversarial perspective. We start with introducing a novel taxonomy, classifying existing methods based on their reliance on internal features (IF) (inherent to the data) versus external features (EF) (artificially introduced for auditing). Subsequently, we formulate two attack types: \textit{evasion attacks}, designed to conceal the use of a dataset, and \textit{forgery attacks}, intending to falsely implicate an unused dataset. Building on the understanding of existing methods and attack objectives, we further propose systematic attack strategies: decoupling, removal, and detection for evasion; adversarial example-based methods for forgery. These formulations and strategies lead to our new benchmark, \sys, comprising 17 evasion attacks, 5 forgery attacks, and 9 representative auditing methods. Extensive evaluations using \sys reveal that none of the evaluated auditing methods are sufficiently robust or distinctive under adversarial settings. These findings underscore the urgent need for developing a more secure and reliable dataset auditing method capable of withstanding sophisticated adversarial manipulation. Code is available in \url{https://github.com/shaoshuo-ss/DATABench}.
}
\keywords{Dataset Auditing, Copyright Protection, Privacy, Evasion Attack, Trustworthy AI}

\maketitle

\section{Introduction}
\label{sec:intro}


The rapid advancement of Deep Learning (DL) has propelled it into widespread applications across various real-world systems, such as computer vision~\citep{he2016deep}, artificial intelligence chatbots~\citep{guo2025deepseek}, and self-driving vehicles~\citep{chen2024end}. Currently, DL models primarily make predictions based on their knowledge learned from a dataset. As such, its success highly relies on and is inseparable from high-quality datasets~\citep{radford2021learning, krauss2024verify}. As a critical resource and fundamental component of DL, datasets largely determine the performance of the models trained on them. However, developers of DL models and applications usually do not disclose the training datasets they use. This lack of transparency gives rise to potential privacy and copyright concerns. For instance, training data may include unauthorized private information. Commercially deployed models may have been trained using copyrighted datasets that prohibit commercial use. Additionally, many existing data protection regulations, such as GDPR, have granted data owners the right to be informed of their data usage. Consequently, auditing the datasets used in DL models has become a critical issue for safeguarding data privacy and copyright~\citep{huang2024general,du2025sok}.

\begin{figure}[t]
    \centering
    \includegraphics[width=0.99\linewidth]{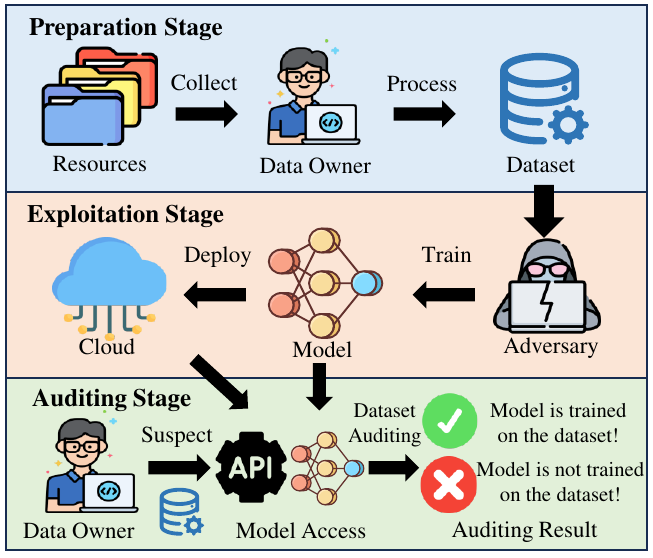}
    \caption{The pipeline of dataset auditing. In the preparation stage, the data owner collects and publishes a dataset. In the exploitation stage, an adversary may copy or steal the dataset and train a model for unauthorized purposes. Finally, in the auditing stage, the data owner audits the suspicious model and verifies whether it is trained on the protected dataset.}
    \label{fig:auditing}
    \vspace{-1em}
\end{figure}

Dataset auditing\footnote{Existing literature may use different terminologies, such as dataset ownership verification~\citep{li2022untargeted}, dataset copyright auditing~\citep{du2025sok}, or data-use auditing~\citep{huang2024general}. In this paper, we standardize the term as \emph{dataset auditing} in DL.} in DL~\citep{li2022untargeted, du2025sok, huang2025instance} refers to determining whether a specific dataset is used to train a suspicious model. Typically, as shown in Figure~\ref{fig:auditing}, there are three stages involved in dataset auditing~\citep{li2023black}. In the first stage, the data owner collects some data from various resources. Subsequently, they process the dataset (\eg, data annotation) to make it suitable for usage or release. Then, in the second stage, an adversary may get access to the dataset through copying or stealing and utilize it to train a DL model without authorization. Third, if a third-party DL model (dubbed a `suspicious model') is suspected of using a protected dataset unauthorizedly, dataset auditing techniques will be adopted to verify it. However, there exist several challenges to achieving a reliable and accurate dataset auditing. \textbf{(1)} It is hard to uncover the traces of a dataset only from the parameters, intermediate results, or even black-box access to a trained DL model. These revealed clues may not be directly associated with the data. \textbf{(2)} The adversary can use various different model architectures and training techniques to produce the model. The diversity and the training-agnostic situation make it difficult to achieve robust and distinctive dataset auditing. \textbf{(3)} The adversary could intentionally process the data or model to compromise the auditing procedure, challenging the robustness of data auditing solutions. As such, achieving reliable dataset auditing is a non-trivial issue in practice~\citep{du2025sok, sablayrolles2020radioactive,chen2024catch}.

Although several existing works have proposed promising solutions to dataset auditing~\citep{tang2023did, guo2024zeromark, maini2020dataset}, there is still a significant gap between how the robustness and distinctiveness of existing works are evaluated (often in controlled environments) and how they perform in real-world practice. Existing dataset auditing research primarily considers \emph{naive} operations and techniques in DL under constrained settings, such as training with models of different architectures, fine-tuning, and pruning, as potential adversarial vectors~\citep{li2022untargeted, du2025sok, guo2023domain}. These works lack the consideration of a `clever' adversary who may design specified and adaptive attacks against dataset auditing. It is challenging to conduct an exhaustive evaluation due to the complexity of training a model in DL and the diversity of attack surfaces. Consequently, the following critical question remains unanswered: \emph{are existing dataset auditing methods reliable under adversarial settings in practice?}

In this paper, we initiate the first comprehensive study on evaluating dataset auditing in DL from an adversarial perspective. Our contributions are as follows.

\vspace{-0.3em}
\begin{itemize}[leftmargin=*]
    \item \textbf{Contribution 1: A Novel Taxonomy of Existing Methods.}
\end{itemize}
\vspace{-0.3em}

We first present the formal definition of dataset auditing. We then revisit existing auditing methods and introduce a novel taxonomy of them. Beyond simply classifying existing methods according to the capabilities (\eg, white/black-box) and operations (\eg, intrusive/non-intrusive) of the data owner, we split the features of a protected dataset into internal features (IF) and external features (EF), and correspondingly classify auditing methods into IF-based and EF-based methods. IF refers to the feature that inherently exists in the data, and the model learns from IF to gain utility. IF-based methods analyze the characteristics of IFs (\eg, overfitting) to achieve auditing. In contrast, EF is the artificial feature introduced by dataset auditing methods and can lead the model to have defender/owner-specified behaviors (\eg, misclassification). EF-based methods achieve auditing by attempting to trigger these behaviors. This IF/EF distinction provides deeper insight into the underlying mechanisms of auditing methods and critically facilitates the systematic design of attacks, which previous taxonomies did not readily support.

\vspace{-0.3em}
\begin{itemize}[leftmargin=*]
    \item \textbf{Contribution 2: Unified Formulations of Attacks against Dataset Auditing.}
\end{itemize}
\vspace{-0.3em}

We formally formulate two fundamental adversarial objectives against dataset auditing, including evasion attacks and forgery attacks. Evasion attacks aim to conceal the usage of a protected dataset. This type of attack maliciously leads the auditing method to incorrectly determine that the victim dataset was not used for training. Conversely, forgery attacks seek to falsely attribute dataset usage. They deceive the auditing method into concluding that an adversary-designated unused dataset was part of the training dataset of an independent victim model. Furthermore, we systematically identify three key phases where adversaries can intervene: the pre-processing, training, and post-training phases. This comprehensive formulation provides a structured understanding of the attack surface in dataset auditing.


\vspace{-0.3em}
\begin{itemize}[leftmargin=*]
    \item \textbf{Contribution 3: Systematic Attack Strategies.}
\end{itemize}
\vspace{-0.3em}

We propose systematic strategies for crafting effective attacks against dataset auditing, building upon our IF/EF taxonomy and attack formulations. For evasion, we introduce three core strategies according to targeted feature type: \textbf{(1) Decoupling:} Weakening the model's reliance on or the detectability of IFs from the training dataset. \textbf{(2) Removal:} Eliminating the artificially introduced EFs or their corresponding designated behaviors. \textbf{(3) Detection:} Identifying inputs likely to trigger EF-based auditing mechanisms and circumventing the response. For forgery, we propose a unified strategy leveraging adversarial example techniques to mimic the specific signals (IF-based confidence responses or EF-based designated behaviors) expected by the target auditing method. These attack strategies provide concrete guidance for designing attacks and comprehensively evaluating dataset auditing.


\vspace{-0.3em}
\begin{itemize}[leftmargin=*]
    \item \textbf{Contribution 4: Dataset Auditing Attack Benchmark and Extensive Evaluation on Existing Methods.}
\end{itemize}
\vspace{-0.3em}

We develop the \underline{D}ataset \underline{A}udi\underline{T}ing \underline{A}ttack \underline{Bench}mark, \sys, based on the proposed attack formulations and strategies. Specifically, we implement 17 evasion attacks and 5 forgery attacks, and evaluate 9 representative dataset auditing methods on the proposed benchmark. The empirical results reveal that none of the evaluated auditing methods exhibit sufficient robustness or distinctiveness under adversarial settings. We surprisingly find that even simple attacks can significantly compromise the effectiveness of existing methods. The results highlight an urgent need for the development of a more reliable dataset auditing method. 

\sys also provides an extensible open-source toolbox to facilitate future research and development in this crucial area. This toolbox is designed to streamline the process to easily develop new dataset auditing techniques, implement additional attacks, and systematically evaluate their proposed methods against the comprehensive and standardized attack suite provided within our unified benchmark.

\section{Background}
\label{sec:background}

\subsection{Deep Learning} 
\label{sec:dl}


Deep Learning (DL) has become the most famous and widely-used technique to develop an artificial intelligence (AI) application during the last decade~\citep{he2016deep, radford2021learning}, especially with the growth of large foundation models~\citep{guo2025deepseek}. DL primarily leverages deep neural networks (DNNs) as the core model architecture to achieve a wide range of functions.

During the development of a DL model $\mathcal{M}$, the model developer needs to first collect a dataset $\mathcal{D}=\{(\bm{x}_i, \bm{y}_i)\}_{i=1}^{|\mathcal{D}|}$. The dataset $\mathcal{D}$ is related to the developer-specified DL task. The dataset consists of various input samples $\bm{x}_i$ and their corresponding label $\bm{y}_i$. 
Subsequently, the developer designs a loss function $\mathcal{L}(\cdot)$ and optimizes the model $\mathcal{M}$ according to the following optimization problem:
\begin{equation}
    \label{eq:dl}
    \widehat{\mathcal{M}} = \arg\min_\mathcal{M} \sum_{i=1}^k\mathcal{L}(\mathcal{M}(\bm{x}_i), \bm{y}_i).
\end{equation}

Currently, most of the DL techniques and models are data-centric and their developments highly rely on a high-quality dataset~\citep{radford2021learning}. This reliance, however, raises significant concerns regarding data privacy and copyright. For instance, many DL models are trained on large-scale datasets that may contain sensitive or proprietary information, leading to potential violations of user privacy and intellectual property rights~\citep{ carlini2022membership, du2025sok}. Consequently, data protection and building high-quality datasets are equally important in practice.

\subsection{Data Protection}

Data protection has long been a critical aspect of information management~\citep{meurisch2021data, li2025rethinking, luo2026shadow, zhu2025stealing}. 
In the last several decades, researchers have developed many different techniques to achieve various requirements of data protection, as follows.

\partitle{Encryption} Encryption-based techniques aim to transform the sensitive data into unreadable ciphertext, and any unauthorized party without the secret key cannot access the data~\citep{bhanot2015review, fiege1987zero}. Advancements in encryption-based techniques like homomorphic encryption (HE)~\citep{paillier1999public, chielle2025recurrent} allow computations on encrypted data, making it promising to be adopted in DL systems. However, encryption-based methods suffer from the limitations of excessive latency and resource demands. 

\partitle{Differential Privacy (DP)} DP ensures that the inclusion or exclusion of any single data point in a dataset does not significantly affect the output~\citep{dwork2006calibrating}. This is typically achieved by injecting controlled noise into data, queries, or model training processes~\citep{dwork2006calibrating, abadi2016deep}. While DP provides strong, mathematically provable privacy guarantees, its application in DL introduces significant challenges. For instance, the added noise often degrades model utility. The trade-off between privacy and utility makes DP a powerful but costly solution for privacy-preserving DL~\citep{abadi2016deep, yu2019differentially}.

\partitle{Traditional Data Watermarking} Traditional data watermarking techniques help track ownership and usage of data by embedding hidden markers (\ie, watermark) into the data~\citep{hussain2022faststamp,zhang2025icmarks}. The data owner can extract the watermark from the data and prove their ownership. Multiple solutions to embedding a watermark into data have been proposed, such as embedding watermarks into the frequency domain or utilizing DL models~\citep{cayre2005watermarking, baluja2017hiding}. 
Recently, some works have also explored using data watermarking to mark AI-generated content and models~\citep{zhang2024remark, shao2025explanation, krauss2024clearstamp, ren2024sok}. 
However, traditional data watermarking techniques may not be feasible for dataset auditing since the watermark in the data can hardly be inherited by the DL models and their owners will not disclose their training details (especially their training data).

\partitle{Unlearnable Data} This technique aims to add perturbations to the data so that the DL model cannot learn from it during training~\citep{huang2021unlearnable, wang2025provably}. The model trained on the unlearnable dataset will have a low utility. However, in the scenario of protecting a public dataset where we intend to prevent unauthorized usage instead of making it completely useless, these techniques may not be feasible.

In general, previous data protection techniques either have a significant impact on the data and model utility (\eg, HE, DP, and unlearnable data) or are not feasible (\eg, traditional data watermarking) in DL scenarios~\citep{li2022untargeted, guo2023domain}. Consequently, dataset auditing, as a technique that has negligible influence on the data utility and can protect the privacy and copyright of data owners, has become an emerging yet important and promising solution to data protection~\citep{li2025towards, du2025sok}.

\section{Dataset Auditing in DL} 
\label{sec:auditing}




\subsection{Towards a Formal Definition of Dataset Auditing}
\label{sec:define}

Before defining dataset auditing, we hereby first introduce the general threat model of dataset auditing in DL~\citep{li2022untargeted, du2025sok}. Specifically, there are two parties involved in the threat model, the \emph{data owner} and the \emph{adversary}\footnote{Some existing literature~\citep{liu2024false, shao2025fit, du2025sok} also introduces a trustworthy third-party \emph{auditor} (or \emph{verifier}) in their threat model to conduct the audit. However, since the auditor does not affect the technical design of dataset auditing methods, we exclude it to avoid potential confusion.}. The data owner wants to collect, store, and publish a DL dataset. However, there exists the risk that an adversary may unlawfully copy or steal the dataset and train their own model for unauthorized usage. Such a misbehavior may compromise the privacy or intellectual property rights of the data owner. As such, the data owner seeks an effective dataset auditing method to confirm whether the suspicious model of the adversary is trained on their dataset. We formally define dataset auditing as follows.

\begin{definition}[Dataset Auditing]
\label{def:audit}
    Given a dataset $\mathcal{D}$, a suspicious model $\mathcal{M}(\cdot)$, and some auxiliary information $\mathcal{I}$, a dataset auditing method $\mathcal{A}(\cdot)$ outputs $1$ if the model is (partly) trained on the dataset $\mathcal{D}$, and 0 otherwise:
    \begin{equation}
        \mathcal{A}(\mathcal{D}, \mathcal{M}, \mathcal{I})= 
        \left\{\begin{aligned}
            1,\ &\mathcal{M}\ \text{is trained on}\ \mathcal{D} \\
            0,\ &\mathcal{M}\ \text{is not trained on}\ \mathcal{D} \\
        \end{aligned}
        \right..
    \end{equation}
\end{definition}

The auxiliary information $\mathcal{I}$ in Definition~\ref{def:audit} varies depending on the specific dataset auditing method. Common forms of $\mathcal{I}$ include auxiliary datasets used for auditing~\citep{shokri2017membership}, specific trigger patterns~\citep{li2022untargeted}, and watermark information~\citep{huang2024general}.

Notably, if the number of samples in $\mathcal{D}$ is $1$ (\ie, $|\mathcal{D}|=1$), the definition of dataset auditing is similar to membership inference~\citep{carlini2022membership, peng2025diffence, shang2025defending}. As such, existing membership inference techniques can be easily adapted to achieve dataset auditing, which will be further discussed in Section~\ref{sec:technique}. 

\partitle{Main Pipeline of Dataset Auditing} In general, there are three stages in a typical pipeline of dataset auditing.

\begin{enumerate}
    \item \textbf{Preparation Stage.} The data owner collects the data and organizes it into a raw dataset $\mathcal{D}$. Then, before publishing the dataset, the data owner can also apply a processing function $\mathcal{P}(\cdot)$, as follows.
    \begin{equation}
        \widehat{\mathcal{D}} = \mathcal{P}(\mathcal{D}, \mathcal{I}),
    \end{equation}
    where $\mathcal{I}$ is the auxiliary information needed in the processing function. $\mathcal{P}(\cdot)$ may be used to add specific perturbations or watermarks to the dataset, or just keep $\widehat{\mathcal{D}}=\mathcal{D}$. Then, the data owner publishes the dataset $\widehat{\mathcal{D}}$.
    \item \textbf{Exploitation Stage.} After the data owner publishes the dataset $\widehat{\mathcal{D}}$, an adversary may steal or copy $\widehat{\mathcal{D}}$ to train their own model $\mathcal{M}$ for unauthorized purposes. They can adopt any training techniques and conduct certain operations to intentionally compromise the dataset auditing mechanism. 
    \item \textbf{Auditing Stage.} If the data owner finds a model\footnote{Following prior works~\citep{du2025sok, li2022untargeted, huang2024general, li2023black}, we do not focus on how to identify these `suspicious models', which may be discovered through non-technical ways. For instance, if a model's task closely aligns with that of a dataset, the model may be regarded as a `suspicious model'.} that is suspected of unauthorizedly using the dataset $\widehat{\mathcal{D}}$, they adopt a dataset auditing method $\mathcal{A}$ to judge whether the misbehavior exists. If $\mathcal{A}(\widehat{\mathcal{D}}, \mathcal{M}, \mathcal{I})$ outputs $1$, the adversary can be accused of unauthorizedly using the dataset and infringing the copyright or privacy.
\end{enumerate}

\firstpartitle{Assumptions of Data Owner} The data owner intends to publish a dataset and audit any unauthorized usage of it. The capabilities of the data owner are as follows.
\begin{itemize}
    \item In the preparation stage, the data owner has full control of this stage and the dataset. They can determine what preprocessing steps $\mathcal{P}(\cdot)$ to apply to the dataset and which portions of the data to release.
    \item In the exploitation stage, the data owner has no knowledge of the operations of the adversary. They also cannot interfere with the training of the adversary's model.
    \item In the auditing stage, the data owner only has access to the adversary's trained model. 
\end{itemize}

According to the operation and capability of the data owner, existing dataset auditing methods can be further classified according to the following simple taxonomies~\citep{du2025sok}.

\begin{definition}[Intrusive and Non-intrusive Dataset Auditing]
\label{def:intrusive}
    In the preparation stage, if the data owner does not need to alter the raw dataset, \ie, $\mathcal{D}=\widehat{\mathcal{D}}$, the dataset auditing method is a non-intrusive method, otherwise an intrusive one.
\end{definition}

\begin{definition}[White-box and Black-box Dataset Auditing]
\label{def:box}
    In the auditing stage, if $\mathcal{A}$ only needs API access to the suspicious model (\ie, only get the outputs of given inputs), it is defined as a black-box method. Otherwise, if additional information about the suspicious model is needed (\eg, the parameters or the intermediate results during inference), $\mathcal{A}$ is a white-box method.
\end{definition}

\subsection{Dataset Auditing Taxonomy}
\label{sec:technique}

Existing works~\citep{du2025sok} primarily introduce simple taxonomies such as Definitions~\ref{def:intrusive} or \ref{def:box}. However, these taxonomies offer limited aid in understanding the underlying insights of dataset auditing methods and designing effective attack strategies to evaluate their reliability. As such, we propose a new taxonomy based on the origin of the features exploited during the auditing process. Specifically, we split the data features into internal and external features and classify existing dataset auditing methods into the following two main categories.




\partitle{Internal Feature-based Dataset Auditing} This class of methods relies on the internal feature (IF) of the dataset. IFs refer to features that naturally exist within the data itself. Learning from these features builds up the utility of the model on the primitive task. The core insight of IF-based methods is to exploit the characteristics of the IFs (\eg, overfitting) during or after training.

Dataset auditing methods based on membership inference attacks (MIA)~\citep{shokri2017membership, he2024difficulty, liu2022your} fall into this category. MIA aims to determine whether a given data instance was included in a model's training dataset. Given a trained model and a specific data point, MIA outputs a confidence score that reflects the likelihood that the instance was part of the training set. Although originally proposed as a privacy attack, MIA can be readily adapted to the task of dataset auditing by aggregating the confidence scores across all the samples in a dataset (\eg, calculating the average confidence score).

Additionally, some existing works explore auxiliary mechanisms to enhance the effectiveness of MIA in the context of dataset auditing. Some studies propose Dataset Inference (DI)~\citep{maini2020dataset, dziedzic2022dataset}, which treats the dataset as a whole and utilizes the prediction margin to conduct dataset auditing. \citep{maini2024llm} also explores aggregating results of different MIAs. Data-use Auditing (DUA)~\citep{huang2024general} adds perturbations to craft two datasets that are significantly different from each other in the feature space. DUA achieves dataset auditing by testing whether the confidence of being a member on one dataset is significantly larger than the other. These mechanisms make the IF-based dataset auditing more feasible in practice.



\partitle{External Feature-based Dataset Auditing} In contrast, external feature (EF)-based methods introduce external or artificial features into the dataset. 
The model trained on the protected dataset with external features will have identifiable behavior that is hardly learned from other datasets. This class of methods achieves dataset auditing by verifying whether the suspicious model can trigger the designated behavior. 

Existing methods have explored different `identifiable behaviors' for auditing. Radioactive Data~\citep{sablayrolles2020radioactive, atli2022effectiveness} induces distributional shifts in feature space. Misclassification on specific samples (\ie, backdoor behaviors) is also widely used~\citep{du2025sok}. A line of research utilizes backdoor techniques~\citep{shan2022poison, li2023black, li2022untargeted, tang2023did, guo2024zeromark} to mark the dataset. Any model trained on a marked dataset will misclassify samples that contain specific trigger patterns. In contrast, Domain Watermark (DW)~\citep{guo2023domain} introduces EFs that enable the model to correctly classify certain hard samples with high accuracy. DW does not deliberately induce misclassification and is therefore inherently harmless.

\partitle{Differences with Prior Taxonomy} The intrusive and non-intrusive taxonomy describes the owner-side operations, whereas our IF/EF taxonomy is defined by the inherent nature of auditing signals. This distinction is critical because it directly shapes the design space and feasibility of attack strategies (details in Section~\ref{sec:attacks}), given that IFs cannot be completely removed by adversaries. For example, DUA~\citep{huang2024general} highlights the limitations of the prior taxonomy. It does not introduce any identifiable EF signal; instead, it amplifies the discriminability of IFs to enhance membership evidence through perturbation injection. Therefore, although DUA is categorized as intrusive, it remains fundamentally IF-based.


\subsection{Design Objectives}
\label{sec:objective}

The objectives of designing a feasible dataset auditing method can be summarized as follows:

\begin{itemize}[leftmargin=*, itemsep=0.3em]
    \item \textbf{Effectiveness:} The dataset auditing should precisely identify the usage of the dataset from the suspicious model.
    \item \textbf{Fidelity:} The dataset auditing method should have a negligible impact on the normal utility of the dataset, \ie, the model trained on the published dataset $\widehat{\mathcal{D}}$ and the raw dataset $\mathcal{D}$ should have minimum difference regarding the primitive task of the dataset.
    \item \textbf{Robustness:} The dataset auditing method should not be significantly affected by an adversary's attacks. 
    \item \textbf{Distinctiveness:} The dataset auditing method should not identify other unused datasets as being used for training.
    \item \textbf{Stealthiness:} The perturbations added to the published dataset $\widehat{\mathcal{D}}$ cannot be easily perceived. 
    \item \textbf{Efficiency:} A good dataset auditing method should introduce as little additional time overhead as possible.
\end{itemize}

In this paper, we primarily focus on two objectives, \emph{robustness} and \emph{distinctiveness}, under adversarial settings.


\section{Attack Methodology}
\label{sec:attacks}

As introduced in Section~\ref{sec:objective}, robustness and distinctiveness are two fundamental properties of a practical dataset auditing method. However, there is still a significant gap between the constrained settings of existing works and practice for evaluating these two properties. Existing dataset auditing research~\citep{li2022untargeted, du2025sok, guo2023domain} primarily considers common yet simple operations in DL and lacks dedicated attack evaluation. To bridge this gap, we propose a unified and comprehensive formulation and methodology of attacks against dataset auditing. We also present systematic attack strategies and implementations of these attacks. All these elements lead to the development of our Dataset Auditing Attack Benchmark, \sys.




\subsection{Attack Formulation}

In our threat model, the data owner aims to detect any unauthorized usage of the dataset via dataset auditing. However, an adversary may conduct attacks to obfuscate the results of dataset auditing. In this section, we present two different types of attacks, \emph{evasion attack} and \emph{forgery attack}, based on the attack objectives. Intuitively, an evasion attack is to make a `trained' dataset be mistakenly identified as `non-trained'. In contrast, a forgery attack is to make a `non-trained' dataset be identified as `trained'. Their formal definitions are as follows.

\begin{definition}[Evasion Attack against Dataset Auditing]
    An evasion attack seeks to conceal the use of a victim dataset. Its goal is to make the dataset auditing method mistakenly infer that the victim dataset was not used during training. Formally, given a dataset $\mathcal{D}$ and any dataset auditing method $\mathcal{A}(\cdot)$, an evasion attack aims to find an approach $\mathcal{E}$ to produce a model $\widetilde{\mathcal{M}}$ trained on $\mathcal{D}$ and have
    \begin{equation}
        \mathcal{A}(\mathcal{D}, \widetilde{\mathcal{M}}, \mathcal{I})=0,\  \widetilde{\mathcal{M}}=\mathcal{E}(\mathcal{D}),
    \end{equation}
    where $\mathcal{I}$ is the auxiliary information required in the process of dataset auditing. The output $0$ indicates that $\widetilde{\mathcal{M}}$ is not trained on $\mathcal{D}$. Meanwhile, $\widetilde{\mathcal{M}}$ needs to have a similar utility with the model directly trained on $\mathcal{D}$.
\end{definition}

\partitle{Adversary's Assumptions in Evasion Attacks} In the evasion attack, the adversary has full control of the model and dataset in the whole training process, including the preprocessing algorithm, the model architecture, and the training method. The adversary also has a small (benign) local dataset related to the model to facilitate evasion attacks.

\begin{definition}[Forgery Attack against Dataset Auditing]
    A forgery attack aims to falsely attribute the usage of a dataset that was not actually used during training. The adversary attempts to carefully construct a malicious dataset $\mathcal{D}_a$ and auxiliary information $\mathcal{I}_a$ such that the dataset auditing method will incorrectly identify $\mathcal{D}_a$ as being in the training set of an independent model $\mathcal{M}_u$, as follows:
    \begin{equation}
        \mathcal{A}(\mathcal{D}_a, \mathcal{M}_u, \mathcal{I}_a)=1,
    \end{equation}
    where $\mathcal{M}_u$ is not actually trained on $\mathcal{D}_a$.
\end{definition}

\partitle{Adversary's Assumptions in Forgery Attacks} In the forgery attack, the adversary has access to a small clean dataset that is related to the task of the target independent model $\mathcal{M}_u$. The adversary also has access to $\mathcal{M}_u$. We take two different levels of access into consideration.
\begin{itemize}[leftmargin=*]
    \item \textbf{White-box Access}: The adversary has full access to the target model, including the parameters and architecture. This scenario may arise when an adversary attempts to attack an open-source model.
    \item \textbf{Black-box Access}: The adversary has API access to the target model, \ie, the inputs and the outputs of the model. This scenario may arise when an adversary tries to attack a model deployed as a service.
\end{itemize}








\subsection{Evasion Attacks}

\begin{figure}
    \centering
    \includegraphics[width=0.99\linewidth]{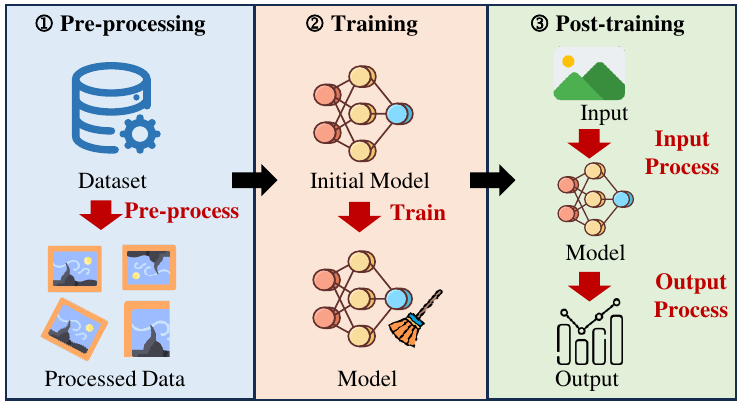}
    \caption{The attack phases in attacking dataset auditing. In the pre-processing phase, the adversary can apply a pre-processing method to the dataset. In the training phase, the adversary utilizes any training technique to get a well-performed and auditing-free model. In the post-training phase, the adversary can alter the inference process of the deployed model to obfuscate dataset auditing.}
    \label{fig:evasion}
    \vspace{-1em}
\end{figure}

\partitle{Attack Phases} In evasion attacks, the adversary has full control of the training process and can conduct any operation trying to evade dataset auditing. We split the model training process into three different phases, pre-processing, training, and post-training, as shown in Figure~\ref{fig:evasion}.

\begin{enumerate}
    \item \textbf{Pre-processing Phase:} In this phase, the adversary can apply any pre-process method to the \emph{dataset} before training to evade dataset auditing.
    \item \textbf{Training Phase:} In this phase, the adversary aims to apply different methods to train a high-performance \emph{model} and, in the meantime, evade dataset auditing.
    \item \textbf{Post-training Phase:} In this phase, we assume that the model has been trained and the parameters of the model are fixed. However, the adversary can change the \emph{procedure of model inference} to evade dataset auditing.
\end{enumerate}

\partitle{Attack Strategy} Based on Section~\ref{sec:technique}, existing dataset auditing methods can be classified into internal feature (IF) based methods and external feature (EF) based methods. This taxonomy can facilitate the design of attacks, as follows.
\begin{enumerate}
    \item \textbf{Attack Strategy against IF-based Methods:} Since internal features are important and indispensable for the utility of the model, the adversary cannot straightforwardly remove them. However, the adversary can undermine the effectiveness of dataset auditing by weakening the association between the model and the training dataset. This strategy is named \textbf{Decoupling}.
    \item \textbf{Attack Strategy against EF-based Methods:} To attack EF-based methods, the external features of a dataset do not have a direct correlation with model utility. They primarily lead to the model's specific behaviors when inputting specific samples. As such, two distinct strategies can be employed to evade EF-based dataset auditing.
    \begin{enumerate}
        \item \textbf{Removal:} This type of attack intends to remove the external features from the dataset and the designated behaviors incurred by the dataset auditing method. 
        \item \textbf{Detection:} This type of attack aims to detect those specific samples that can trigger the designated unique behaviors of the model and return a random output for them\footnote{We consider this type of attack primarily in the post-training phase. This is because it arguably achieves data protection if the adversary detects and removes the marked samples before deploying the model (\ie, the model is not trained on the protected data).}.
    \end{enumerate}
\end{enumerate}



\firstpartitle{Attack Implementations} Based on the attack strategies, we can design various attacks in different attack phases. The implementations of evasion attacks in \sys are as follows.

    \partitle{1) Attacks in the Pre-processing Phase}
    \begin{itemize}
        \item \textbf{Decoupling}: To reduce the association between the trained model and the dataset, the adversary can adopt operations to make the samples in the dataset less similar to the original ones. As such, \emph{data augmentation} and \emph{data synthesis} are two approaches to achieve such a goal. Data augmentation applies transformations such as cropping and flipping to increase the diversity of training data~\citep{shorten2019survey}. Data synthesis is an advanced technique that utilizes the capability of generative models~\citep{li2025easy, dockhorn2023differentially}. It first trains a generative model (\eg, a diffusion model~\citep{ho2020denoising}) on the dataset and then uses the model to generate a synthetic dataset. The model is subsequently trained on the synthetic dataset instead of the original one.
        \item \textbf{Removal}: Since the external features usually tend to be `noise' in the dataset, denoising techniques can be adopted to remove them. We consider two classes of denoising methods, \emph{traditional denoising filters} and \emph{DL-based denoising methods}. The former reduces noise by smoothing pixel values~\citep{fan2019brief} and the latter exploits DL models (\eg, autoencoder~\citep{li2023comprehensive}) to remove the noise while preserving the important features for model training.
    \end{itemize}
    \partitle{2) Attacks in the Training Phase}
    \begin{itemize}
        \item \textbf{Decoupling:} The adversary can reduce the model's overfitting during training to weaken its association with the training dataset. For instance, \emph{regularization} is a common approach in practice to achieve such a goal. Additionally, during training, the adversary can apply \emph{Differential Privacy (DP) mechanisms}~\citep{dwork2006calibrating} (\eg, DP-SGD~\citep{abadi2016deep}) to make any single data point indistinguishable.
        \item \textbf{Removal:} To eliminate potential abnormal behaviors during training, we consider \emph{adversarial training} and \emph{training-phase backdoor defense} as two representative removal attacks. Adversarial training~\citep{tramer2018ensemble} improves the prediction robustness of the model by including adversarial examples (AE) during the training process. AEs are inputs intentionally perturbed to lead to misclassification; Training-phase backdoor defense~\citep{gao2023backdoor, huang2022backdoor} aims to mitigate hidden external features in the training data by trying not to learn from them. Both ways may remove the designated behaviors and evade dataset auditing.
    \end{itemize}
    \partitle{3) Attacks in the Post-training Phase}
    \begin{itemize}
        \item \textbf{Decoupling:} In this phase, the adversary can also apply DP mechanisms to add noise during the model inference. As such, we implement \emph{noisy output}~\citep{du2025sok} and \emph{noisy feature}~\citep{du2023dp} that perturb the outputs and intermediate results of the model, respectively, during inference to evade dataset auditing.
        \item \textbf{Removal:} Since the model parameters are fixed, the adversary cannot remove the external features from the model. However, it can still hide the external behaviors by changing inference mechanisms. \emph{Randomized smoothing (RS)} and \emph{transformation-based backdoor defense} are two typical methods. Specifically, RS~\citep{cohen2019certified} adds noise to the inputs and calculates the aggregated results of the outputs on noisy inputs; Transformation-based backdoor defense~\citep{li2021backdoor} (\eg, reprogramming~\citep{chen2025refine}) neutralizes the designated behaviors by applying input transformations.
        \item \textbf{Detection:} This type of attack aims to detect the trigger samples used for dataset auditing. Since the trigger samples tend to be different from normal samples, we implement traditional \emph{outlier detection}~\citep{boser1992training} based on the features of inputs and advanced \emph{backdoor sample detection}~\citep{hou2024ibd, guo2023scale}. If these methods identify the input samples as an outlier or backdoor sample, the adversary can refuse to reply to evade dataset auditing.
    \end{itemize}

\subsection{Forgery Attacks}


\partitle{Attack Strategy} The goal of the forgery attack is to craft some samples to falsely claim that an independent model $\mathcal{M}_u$ is trained on these samples. Assuming that the adversary has an auxiliary dataset $\mathcal{D}_a$, he can optimize a set of perturbations $\mathcal{R}=\{\bm{r}_i\}_i^{|\mathcal{D}_a|}$ to make the dataset auditing method output $1$. Ideally, this can be formalized as an optimization problem:
\begin{equation}
\begin{aligned}
    &\text{Find}\ \mathcal{R}\ \text{s.t.}\ \mathcal{A}(\mathcal{D}_a', \mathcal{M}_u, \mathcal{I}_a)=1, \\
    \mathcal{D}_a'&= \left\{(\bm{x}_i + \bm{r}_i, \bm{y}_i)| (\bm{x}_i, \bm{y_i})\in \mathcal{D}_a, \bm{r}_i\in \mathcal{R} \right\}_{i=1}^{|\mathcal{D}_a|},
\end{aligned}
\end{equation}
where $(\bm{x}_i, \bm{y}_i), \bm{r}_i$ are the $i$-th element in $\mathcal{D}_a$ and $\mathcal{R}$. Inspired by \citep{liu2024false, shao2025fit}, the adversary could adopt adversarial attack techniques~\citep{goodfellow2014explaining, ren2020adversarial} to achieve the forgery attack.
\begin{itemize}
    \item IF-based methods primarily rely on the `overfitting' of the internal features to achieve dataset auditing. Therefore, the adversary can increase the output confidence of the model on these forged samples to make them similar to `overfitted training samples'.
    \item EF-based methods depend on abnormal behaviors (typically, misclassification) of the model. As such, the adversary can optimize the perturbations to lead these forged samples to trigger misclassification.
\end{itemize}

\partitle{Attack Implementation} Assuming that $\mathcal{T}(\mathcal{M}, \bm{x}, \bm{y}_t)$ is a targeted adversarial attack to achieve the following goal:
\begin{equation}
\begin{aligned}
    \mathcal{T}(\mathcal{M}, \bm{x}, \bm{y}_t) =& \arg\min_{\bm{r}}\mathcal{L}(\mathcal{M}(\bm{x}+\bm{r}), \bm{y}_t),\\
    \text{s.t.}\ &\|\bm{r}\|\leq\epsilon,
\end{aligned}
\end{equation}
and $\bm{y}$ is the ground-truth label of $\bm{x}$. To attack IF-based methods, the adversary can set $\bm{y}_t=\bm{y}$ to increase the prediction confidence. In contrast, the adversary can set $\bm{y}_t\neq \bm{y}$ to attack EF-based methods. Notably, slight differences exist among the implementations of forged attacks against different dataset auditing methods, but they all follow the same strategy. We include the detailed descriptions in Appendix~\ref{apd:detail}.

Arguably, the adversary can apply any targeted adversarial attacks to achieve such a goal. Specifically, we implement 5 different forgery attacks, as shown in Table~\ref{tab:attackoverview}.


\begin{table*}[t]
    \centering
    \tabcolsep=1.0mm
    \renewcommand{\arraystretch}{1.3}
    \caption{A list of attacks against dataset auditing implemented in \sys. The `Data' column shows whether they need auxiliary data for attacks.}
    \label{tab:attackoverview}
    \scalebox{0.75}{
    \begin{tabular}{ccccc} 
        \toprule 
        Attack & Type & Strategy & Phase & Data \\
        \midrule 
        Data Augmentation & Evasion & Decoupling & Pre-processing & \xmark \\
        Data Synthesis~\citep{ho2020denoising} & Evasion & Decoupling & Pre-processing & \cmark \\
        Gaussian Filter, Median Filter, Wavelet Filter~\citep{fan2019brief} & Evasion & Removal & Pre-processing & \xmark \\
        Autoencoder-based Denoising~\citep{li2023comprehensive} & Evasion & Removal & Pre-processing & \cmark \\
        \midrule 
        Regularization, DP-SGD~\citep{abadi2016deep} & Evasion & Decouping & Training & \xmark \\
        Adversarial Training~\citep{tramer2018ensemble} & Evasion & Removal & Training & \xmark \\
        Adaptively Splitting Poisoned Dataset (ASD)~\citep{gao2023backdoor} & Evasion & Removal & Training & \cmark\\
        \midrule 
        Noisy Output, Noisy Feature~\citep{du2025sok} & Evasion & Decoupling & Post-training & \xmark \\
        Randomized Smoothing (RS)~\citep{cohen2019certified} & Evasion & Removal & Post-training & \xmark \\
        Reprogramming~\citep{chen2025refine} & Evasion & Removal & Post-training & \cmark \\
        SCALE-UP~\citep{guo2023scale} & Evasion & Detection & Post-training & \xmark \\
        KNN-based Outlier Detection, SVM-based Outlier Detection & Evasion & Detection & Post-training & \cmark\\
        \midrule 
        FGSM~\citep{goodfellow2014explaining}, PGD~\citep{madry2018towards}, UAP~\citep{moosavi2017universal}, TIFGSM~\citep{dong2019evading}, VNIFGSM~\citep{wang2021enhancing} & Forgery & / & / & \cmark \\
        \bottomrule 
    \end{tabular}
    }
    \vspace{-0.5em}
\end{table*}

\section{Benchmarking Attacks against Dataset Auditing}
\label{sec:evaluation}

\subsection{Overview of \sys}
\label{sec:overview}

In this section, we present the overview of our dataset auditing attack benchmark \sys. Notably, \sys aims to cover both simple and advanced attack methods. Table~\ref{tab:attackoverview} shows the implemented attacks in \sys. We implement a total of 17 evasion attacks, including 6 attacks in the pre-processing phase, 4 attacks in the training phase, and 7 attacks in the post-training phase. Naive attacks such as pruning and quantization are not included since a wide range of papers have demonstrated they are ineffective~\citep{li2022untargeted, du2025sok, huang2024general, li2023black}. Besides, we implement 5 different forgery attacks. Detailed descriptions of these attacks are in Appendix~\ref{apd:detailevasion}. 

In \sys, we implement 9 representative dataset auditing methods, including 4 IF-based methods and 5 EF-based methods. For IF-based methods, besides Dataset Inference (DI)~\citep{maini2020dataset} and Data-use Auditing (DUA)~\citep{huang2024general}, which are designed for dataset auditing, we also adopt two representative MIAs as auditing methods. We select a classic MIA proposed by Shokri et al.~\citep{shokri2017membership} (denoted as MIA) and the state-of-the-art (SOTA) MIA named Rapid~\citep{he2024difficulty}. For EF-based methods, we evaluate DVBW~\citep{li2023black}, UBW (including UBW-P and UBW-C)~\citep{li2022untargeted}, ZeroMark~\citep{guo2024zeromark}, and Domain Watermark (DW)~\citep{guo2023domain}. The detailed settings of these dataset auditing methods can be found in Appendix~\ref{apd:detailaudit}.

\begin{table*}[t]
    \centering
    \tabcolsep=1.5mm
    \renewcommand{\arraystretch}{1.1}
    \caption{Results of evasion attacks in the preprocessing phase. We highlight cells corresponding to failed audits in \failurecolor and those indicating successful audits in \successcolor. (Better viewed in color)}
    \label{tab:preprocessing}
    \scalebox{0.71}{
    \begin{tabular}{ccccccccc} 
        \toprule 
        Type & Auditing Method & Metric & No Attack & Data Synthesis & GaussianFilter & MedianFilter & WaveletFilter & AutoEncoder \\
        \midrule 
        \multirow{10}{*}{IF-based}
        & \multirow{2}{*}{MIA} &Test Acc. (\%, $\uparrow$) & \cellgreen 87.58 & \cellred 78.03 & \cellgreen 87.61 & \cellgreen 80.38 & \cellgreen 76.02 & \cellgreen 85.86 \\
        &  & Score ($\uparrow$) & \cellgreen 0.62 & \cellred 0.47 & \cellgreen 0.62 & \cellgreen 0.55 & \cellred 0.49 & \cellgreen 0.59 \\
        \cmidrule(lr){2-9} 
        & \multirow{2}{*}{Rapid} &Test Acc. (\%, $\uparrow$) & \cellgreen 87.58 & \cellgreen 78.03 & \cellgreen 87.61 & \cellgreen 80.38 & \cellgreen 76.02 & \cellgreen 85.86 \\
        &  & Score ($\uparrow$) & \cellgreen 0.66 & \cellgreen 0.71 & \cellgreen 0.66 & \cellgreen 0.70 & \cellgreen 0.72 & \cellgreen 0.65 \\
        \cmidrule(lr){2-9}
        & \multirow{3}{*}{DI} &Test Acc. (\%, $\uparrow$) & \cellgreen 87.58 & \cellgreen 78.03 & \cellgreen 87.61 & \cellgreen 80.38 & \cellgreen 76.02 & \cellred 85.86 \\
        &  & p-value ($\downarrow$) & \cellgreen $10^{-25}$ & \cellgreen $10^{-3}$ & \cellgreen 0.02 & \cellgreen 0.02 & \cellred 0.36 & \cellred 0.08 \\
        &  & Diff ($\uparrow$) & \cellgreen 1.33 & \cellgreen 0.85 & \cellgreen 0.51 & \cellgreen 0.51 & \cellred 0.00 & \cellred 0.00 \\
        \cmidrule(lr){2-9}
        & \multirow{3}{*}{DUA} &Test Acc. (\%, $\uparrow$) & \cellgreen 87.17 & \cellred 77.35 & \cellgreen 87.58 & \cellgreen 79.31 & \cellgreen 75.82 & \cellgreen 84.76 \\
        &  & Cost (logits) (\%, $\downarrow$) & \cellgreen 16.66 & \cellred 100.00 & \cellgreen 44.11 & \cellgreen 95.66 & \cellgreen 22.11 & \cellgreen 59.60 \\
        &  & Cost (label) (\%, $\downarrow$) & \cellgreen 85.97 & \cellred 100.00 & \cellgreen 75.80 & \cellgreen 99.51 & \cellgreen 22.23 & \cellgreen 84.14 \\
        \midrule 
        \multirow{14}{*}{EF-based}
        & \multirow{3}{*}{DVBW} &Test Acc. (\%, $\uparrow$) & \cellgreen 86.58 & \cellgreen 76.37 & \cellgreen 86.91 & \cellgreen 68.48 & \cellgreen 73.72 & \cellred 83.39 \\
        &  & p-value ($\downarrow$) & \cellgreen 0.00 & \cellgreen 0.00 & \cellgreen 0.00 & \cellgreen 0.00 & \cellred 1.00 & \cellred 1.00 \\
        &  & WSR (\%, $\uparrow$) & \cellgreen 100.00 & \cellgreen 99.96 & \cellgreen 100.00 & \cellgreen 100.00 & \cellred 19.53 & \cellred 16.40 \\
        \cmidrule(lr){2-9}
        & \multirow{3}{*}{UBW-P} & Test Acc. (\%, $\uparrow$) &  \cellgreen 87.28 &  \cellgreen 76.31 & \cellgreen 87.09 & \cellgreen 77.25 & \cellgreen 74.40 & \cellred 84.14 \\
        &  & p-value ($\downarrow$) & \cellgreen 0.00 & \cellgreen $10^{-243}$ & \cellgreen 0.00 & \cellgreen 0.00 & \cellred 1.00 & \cellred 1.00 \\
        &  & WSR (\%, $\uparrow$) & \cellgreen 87.17 & \cellgreen 39.13 & \cellgreen 89.17 & \cellgreen 87.53 & \cellred 29.02 & \cellred 31.16 \\
        \cmidrule(lr){2-9}
        & \multirow{3}{*}{UBW-C} & Test Acc. (\%, $\uparrow$) & \cellgreen 83.05 & \cellred 75.11 & \cellgreen 82.99 & \cellred 75.79 & \cellgreen 72.51 & \cellgreen 81.81 \\
        &  & p-value ($\downarrow$) & \cellgreen $10^{-60}$ & \cellred 1.00 & \cellgreen $10^{-60}$ & \cellred 1.00 & \cellred 1.00 & \cellgreen $10^{-62}$ \\
        &  & WSR (\%, $\uparrow$) & \cellgreen 75.49 & \cellred 27.61 & \cellgreen 76.99 & \cellred 33.33 & \cellred 26.86 & \cellgreen 72.26 \\
        \cmidrule(lr){2-9}
        & \multirow{3}{*}{ZeroMark} &Test Acc. (\%, $\uparrow$) & \cellgreen 87.14 & \cellgreen 75.37 & \cellgreen 86.94 & \cellgreen 70.42 & \cellgreen 71.66 & \cellred 83.83 \\
        &  & p-value ($\downarrow$) & \cellgreen $10^{-137}$ & \cellgreen $ 10^{-126}$ & \cellgreen $10^{-213}$ & \cellred 0.96 & \cellred 1.00 & \cellred 1.00 \\
        &  & WSR (\%, $\uparrow$) & \cellgreen 100.00 & \cellgreen 99.49 & \cellgreen 100.00 & \cellred 100.00 & \cellred 27.56 & \cellred 19.01 \\
        \cmidrule(lr){2-9}
        & \multirow{2}{*}{DW} &Test Acc. (\%, $\uparrow$) &  \cellgreen 86.87 & \cellgreen 74.88 & \cellgreen 87.01 & \cellgreen 77.05 & \cellgreen 72.45 & \cellred 85.05 \\
        &   & WSR (\%, $\uparrow$) & \cellgreen 67.20 & \cellred 3.70 & \cellgreen 33.50 & \cellgreen 85.30 & \cellred 23.00 & \cellred 4.90 \\
        \bottomrule 
    \end{tabular}
    }
    \vspace{-0.5em}
\end{table*}

\begin{table*}[t]
    \centering
    \tabcolsep=1mm
    \renewcommand{\arraystretch}{1.1}
    \caption{Results of evasion attacks in the training phase. We highlight cells corresponding to failed audits in \failurecolor and those indicating successful audits in \successcolor. (Better viewed in color)}
    \label{tab:training}
    \scalebox{0.68}{
    \begin{tabular}{ccccccccc} 
        \toprule 
        Type & Auditing Method & Metric & No Attack & DP-SGD ($\varepsilon=32$) & DP-SGD ($\varepsilon=64$) & AdvTraining-FGSM & AdvTraining-Hybrid & ASD \\
        \midrule 
        \multirow{10}{*}{IF-based} 
        & \multirow{2}{*}{MIA} &Test Acc. (\%, $\uparrow$) & \cellgreen 87.58 & \cellgreen 68.54 & \cellgreen 70.19 & \cellgreen 83.40 & \cellgreen 83.44 & \cellgreen 66.80 \\
         & & Score ($\uparrow$) & \cellgreen 0.62 & \cellred 0.48 & \cellred 0.49 & \cellgreen 0.56 & \cellgreen 0.53 & \cellred 0.20 \\
        \cmidrule(lr){2-9} 
         & \multirow{2}{*}{Rapid} &Test Acc. (\%, $\uparrow$) & \cellgreen 87.58 & \cellgreen 68.54 & \cellgreen 70.19 & \cellgreen 83.40 & \cellgreen 83.44 & \cellgreen 66.80 \\
         & & Score ($\uparrow$) & \cellgreen 0.66 & \cellred 0.40 & \cellred 0.44 & \cellgreen 0.69 & \cellgreen 0.72 & \cellred 0.02 \\
        \cmidrule(lr){2-9}
         & \multirow{3}{*}{DI} &Test Acc. (\%, $\uparrow$) & \cellgreen 87.58 & \cellgreen 68.54 & \cellgreen 70.19 & \cellred 83.40 & \cellred 83.44 & \cellgreen 66.80 \\
         & & p-value ($\downarrow$) & \cellgreen $10^{-25}$ & \cellred 0.06 & \cellgreen $10^{-3}$ & \cellred 0.09  & \cellred 0.18 & \cellred 0.11 \\
         & & Diff ($\uparrow$) & \cellgreen 1.33 & \cellred 0.21 & \cellgreen 0.74 & \cellred 0.19 & \cellred 0.00 & \cellred 0.02 \\
        \cmidrule(lr){2-9}
         & \multirow{3}{*}{DUA} &Test Acc. (\%, $\uparrow$) & \cellgreen 87.17 & \cellgreen 67.63 & \cellgreen 70.45 & \cellgreen 82.98 & \cellgreen 83.75 & \cellred 79.39 \\
         & & Cost (logits) (\%, $\downarrow$) & \cellgreen 16.66 & \cellred 100.00 & \cellred 100.00 & \cellgreen 79.20 & \cellgreen 99.29 & \cellred 100.00 \\
         & & Cost (label) (\%, $\downarrow$) & \cellgreen 85.97 & \cellred 100.00 & \cellred 100.00 & \cellgreen 84.20 & \cellgreen 98.29 & \cellred 100.00 \\
         \midrule 
         \multirow{14}{*}{EF-based}
        & \multirow{3}{*}{DVBW} &Test Acc. (\%, $\uparrow$) & \cellgreen 86.58 & \cellgreen 64.54 & \cellgreen 66.11 & \cellgreen 81.44 & \cellgreen 82.28 & \cellgreen  77.57 \\
         & & p-value ($\downarrow$) & \cellgreen 0.00 & \cellgreen 0.00 & \cellgreen 0.00 & \cellgreen 0.00 & \cellgreen 0.00 &  \cellgreen 0.00 \\
         & & WSR (\%, $\uparrow$) & \cellgreen 100.00 & \cellgreen 100.00 & \cellgreen 100.00 & \cellgreen 100.00 & \cellgreen 100.00 & \cellgreen 78.10 \\
        \cmidrule(lr){2-9}
         & \multirow{3}{*}{UBW-P} & Test Acc. (\%, $\uparrow$) & \cellgreen 87.28 & \cellgreen 64.95 & \cellgreen 66.97 & \cellgreen 81.30 & \cellgreen 82.22 & \cellgreen 68.27 \\
         & & p-value ($\downarrow$) & \cellgreen 0.00 & \cellgreen 0.00 & \cellgreen 0.00 & \cellgreen 0.00 & \cellgreen 0.00 & \cellgreen 0.00 \\
         & & WSR (\%, $\uparrow$) & \cellgreen 87.40 & \cellgreen 83.90 & \cellgreen 86.30 & \cellgreen 87.70 & \cellgreen 76.80 & \cellgreen 90.00 \\
        \cmidrule(lr){2-9}
         & \multirow{3}{*}{UBW-C} & Test Acc. (\%, $\uparrow$) & \cellgreen 83.05 & \cellgreen 66.68 & \cellgreen 68.67 & \cellred 73.22 & \cellgreen 70.76 & \cellgreen 85.38 \\
         & & p-value ($\downarrow$) & \cellgreen $10^{-60}$ & \cellred 1.00 & \cellred 1.00 & \cellred 1.00 & \cellred 1.00 & \cellgreen $10^{-41}$ \\
         & & WSR (\%, $\uparrow$) & \cellgreen 75.49 & \cellred 34.80 & \cellred 35.20 & \cellred 31.10 & \cellred 28.20 & \cellgreen 44.60 \\
        \cmidrule(lr){2-9}
         & \multirow{3}{*}{ZeroMark} &Test Acc. (\%, $\uparrow$) & \cellgreen 87.14 & \cellgreen 67.72 & \cellgreen 69.32 & \cellred 82.17 & \cellred 82.68 & \cellred 78.57 \\
         & & p-value ($\downarrow$) & \cellgreen $10^{-137}$ & \cellred 1.00 & \cellred 1.00 & \cellred 1.00 & \cellred 1.00 & \cellred 1.00 \\
         & & WSR (\%, $\uparrow$) & \cellgreen 100.00 & \cellred 100.00 & \cellred 100.00 & \cellred 95.70 & \cellred 100.00 & \cellred 99.67 \\
        \cmidrule(lr){2-9}
         & \multirow{2}{*}{DW} &Test Acc. (\%, $\uparrow$) & \cellgreen 86.87 & \cellgreen 66.12 & \cellgreen 67.81 & \cellred 85.09 & \cellred 83.91 & \cellgreen 58.65 \\
         & & WSR (\%, $\uparrow$) & \cellgreen 67.20 & \cellred 0.00 & \cellred 0.20 & \cellred 2.40 & \cellred 1.20 & \cellred 0.00 \\
        \bottomrule 
    \end{tabular}
    }
\end{table*}

\begin{table*}[t]
    \centering
    \tabcolsep=1mm
    \renewcommand{\arraystretch}{1.09}
    \caption{Results of evasion attacks in the post-training phase. We highlight cells corresponding to failed audits in \failurecolor and those indicating successful audits in \successcolor. (Better viewed in color)}
    \label{tab:post}
    \scalebox{0.69}{
    \begin{tabular}{ccccccccccc} 
        \toprule 
        Type & Auditing Method & Metric & No Attack & NoisyOutput& NoisyFeature & RS & Reprogramming & SCALE-UP & OD-KNN & OD-SVM \\
        \midrule 
        \multirow{10}{*}{IF-based} 
         & \multirow{2}{*}{MIA} &Test Acc. (\%, $\uparrow$) & \cellgreen 87.58 & \cellgreen 87.26 & \cellgreen 86.88 & \cellred 81.58 & \cellred 85.21 & \cellred 79.32 & \cellgreen 84.08 & \cellgreen 86.57 \\
         & & Score ($\uparrow$) & \cellgreen 0.62 & \cellgreen 0.61 & \cellgreen 0.61 & \cellred 0.00 & \cellred 0.00 & \cellred 0.50 & \cellgreen 0.61 & \cellgreen 0.62 \\
        \cmidrule(lr){2-11} 
         & \multirow{2}{*}{Rapid} &Test Acc. (\%, $\uparrow$) & \cellgreen 87.58 & \cellgreen 87.21 & \cellgreen 87.30 & \cellred 81.58 & \cellred 85.40 & \cellgreen 79.32 & \cellgreen 84.08 & \cellgreen 86.55 \\
         & & Score ($\uparrow$) & \cellgreen 0.66 & \cellgreen 0.65 & \cellgreen 0.65 & \cellred 0.19 & \cellred 0.25 & \cellgreen 0.72 & \cellgreen 0.70 & \cellgreen 0.68 \\
         \cmidrule(lr){2-11}
         & \multirow{3}{*}{DI} &Test Acc. (\%, $\uparrow$) & \cellgreen 87.58 & \cellgreen 87.26 & \cellgreen 86.88 & \cellgreen 81.58 & \cellred 85.69 & \cellgreen 79.32 & \cellred 84.08 & \cellgreen 86.55 \\
         & & p-value ($\downarrow$) & \cellgreen $10^{-25}$ & \cellgreen $10^{-5}$ & \cellgreen $10^{-9}$ & \cellgreen $10^{-7}$ & \cellred 0.09 & \cellgreen $10^{-4}$ & \cellred 0.36 & \cellgreen $10^{-3}$ \\
         & & Diff ($\uparrow$) & \cellgreen 1.33 & \cellgreen 1.15 & \cellgreen 1.25 & \cellgreen 1.27 & \cellred 0.24 & \cellgreen 0.88 & \cellred 0.00 & \cellgreen 0.70 \\
         \cmidrule(lr){2-11}
         & \multirow{3}{*}{DUA} &Test Acc. (\%, $\uparrow$) & \cellgreen 87.17 & \cellgreen 87.12 & \cellgreen 86.93 & \cellgreen 85.58 & \cellgreen 85.59 & \cellgreen 78.46 & \cellgreen 84.01 & \cellgreen 86.08 \\
         & & Cost (logits) (\%, $\downarrow$) & \cellgreen 16.66 & \cellgreen 31.23 & \cellgreen 29.97 & \cellgreen 0.91 & \cellgreen 51.63 & \cellgreen 48.37 & \cellgreen 56.20 & \cellgreen 37.57 \\
         & & Cost (label) (\%, $\downarrow$) & \cellgreen 85.97 & \cellgreen 84.94 & \cellgreen 70.31 & \cellgreen 45.77 & \cellgreen 73.03 & \cellgreen 63.29 & \cellgreen 20.43 & \cellgreen 38.26 \\
         \midrule 
        \multirow{14}{*}{EF-based}
        & \multirow{3}{*}{DVBW} &Test Acc. (\%, $\uparrow$) & \cellgreen 86.58 & \cellgreen 86.23 & \cellgreen 86.13 & \cellred 78.20 & \cellred 83.54 & \cellgreen 77.91 & \cellgreen 83.16 & \cellred 85.74 \\
         & & p-value ($\downarrow$) & \cellgreen 0.00 & \cellgreen 0.00 & \cellgreen 0.00 & \cellred 1.00 & \cellred 1.00 & \cellgreen 0.00 & \cellgreen 0.00 & \cellred 1.00 \\
         & & WSR (\%, $\uparrow$) & \cellgreen 100.00 & \cellgreen 100.00 & \cellgreen 100.00 & \cellred 100.00 & \cellred 9.50 & \cellgreen 81.00 & \cellgreen 99.63 & \cellred 1.79 \\
        \cmidrule(lr){2-11}
         & \multirow{3}{*}{UBW-P} & Test Acc. (\%, $\uparrow$) & \cellgreen 87.28 & \cellgreen 86.98 & \cellgreen 86.98 & \cellred 79.43 & \cellgreen 75.83 & \cellgreen 79.40 & \cellgreen 83.90 & \cellgreen 86.55 \\
         & & p-value ($\downarrow$) & \cellgreen 0.00 & \cellgreen 0.00 & \cellgreen 0.00 & \cellred 1.00 & \cellred 1.00 & \cellgreen 0.00 & \cellgreen 0.00 & \cellgreen 0.00 \\
         & & WSR (\%, $\uparrow$) & \cellgreen 87.39 & \cellgreen 89.05 & \cellgreen 89.73 & \cellred 88.29 & \cellred 25.62 & \cellgreen 87.55 & \cellgreen 90.00 & \cellgreen 87.39 \\
        \cmidrule(lr){2-11}
         & \multirow{3}{*}{UBW-C} & Test Acc. (\%, $\uparrow$) & \cellgreen 83.05 & \cellgreen 82.65 & \cellgreen 86.32 & \cellred 74.93 & \cellred 84.76 & \cellred 77.49 & \cellgreen 80.17 & \cellgreen 82.32 \\
         & & p-value ($\downarrow$) & \cellgreen $10^{-60}$ & \cellgreen $10^{-97}$ & \cellgreen $10^{-97}$ & \cellred 1.00 & \cellred 1.00 & \cellred 1.00 & \cellgreen $10^{-82}$ & \cellgreen $10^{-89}$ \\
         & & WSR (\%, $\uparrow$) & \cellgreen 75.49 & \cellgreen 77.36 & \cellgreen 75.74 & \cellred 98.63 & \cellred 76.36 & \cellred 71.76 & \cellgreen 37.68 & \cellgreen 77.11 \\
        \cmidrule(lr){2-11}
         & \multirow{3}{*}{ZeroMark} &Test Acc. (\%, $\uparrow$) & \cellgreen 87.14 & \cellred 86.82 & \cellred 86.63 & \cellgreen 70.45 & \cellred 84.38 & \cellgreen 76.94 & \cellgreen 83.97 & \cellgreen 86.19 \\
         & & p-value ($\downarrow$) & \cellgreen $10^{-137}$ & \cellred 1.00 & \cellred 1.00 & \cellred 1.00 & \cellred 1.00 & \cellgreen $10^{-143}$ & \cellgreen $10^{-130}$ & \cellgreen $10^{-123}$ \\
         & & WSR (\%, $\uparrow$) & \cellgreen 100.00 & \cellred 100.00 & \cellred 100.00 & \cellred 100.00 & \cellred 10.66 & \cellgreen 100.00 & \cellgreen 100.00 & \cellgreen 100.00 \\
        \cmidrule(lr){2-11}
         & \multirow{2}{*}{DW} &Test Acc. (\%, $\uparrow$) & \cellgreen 86.87 & \cellgreen 86.79 & \cellgreen 86.55 & \cellgreen 82.99 & \cellgreen 62.97 & \cellred 79.36 & \cellred 82.04 & \cellgreen 86.08 \\
         & & WSR (\%, $\uparrow$) & \cellgreen 67.20 & \cellgreen 61.10 & \cellgreen 60.60 & \cellgreen 62.90 & \cellred 11.50 & \cellred 5.60 & \cellred 13.60 & \cellgreen 62.40 \\
        \bottomrule 
    \end{tabular}
    }
    \vspace{-0.5em}
\end{table*}

\subsection{Experimental Settings}

\partitle{Models and Datasets} In the main experiments, we adopt the widely-used CIFAR-10~\citep{krizhevsky2009learning} and ResNet-18~\citep{he2016deep}. The reason for using this simple model and dataset is that almost all the existing dataset copyright auditing methods perform the best in a simple dataset~\citep{du2025sok}. We also include experiments on other models and datasets such as Vision Transformer~\citep{mehta2021mobilevit} and ImageNet~\citep{deng2009imagenet} in Section~\ref{sec:evasion}. We further split the training data: 70\% is the public release dataset (protected), 15\% is the defender's unpublished audit auxiliary dataset, and 15\% is the adversary's unpublished attack auxiliary dataset. Both auxiliary sets are assumed to be trusted and clean (unwatermarked). We train the model both from scratch on the victim dataset and by fine-tuning (Section~\ref{apd:finetuning}). 

\partitle{Attack Settings} In our experiments, we evaluate existing dataset auditing methods on all the attacks in \sys. Given the significant impact of data augmentation and regularization on model utility, we adopt them as default settings. Besides, we consider two variants for adversarial training attacks. `AdvTraining-FGSM' simply uses FGSM~\citep{goodfellow2014explaining} to generate the adversarial examples for training, and `AdvTraining-Hybrid' composes three different adversarial attacks~\citep{goodfellow2014explaining, madry2018towards, carlini2017towards}. We also test two different privacy budgets ($\varepsilon=32$ or $64$) in DP-SGD to simulate different levels of privacy protection. Detailed descriptions and settings can be found in Appendix~\ref{apd:detailevasion}.

\partitle{Dataset Auditing Settings} In our experiments, we consider the ideal situation for dataset auditing. If a dataset auditing method needs to train auxiliary models, we assume that the data owner uses the same model architecture as the adversary. It ensures that we evaluate the worst scenario for the adversary. Detailed settings can be found in Appendix~\ref{apd:detailaudit}.

\partitle{Metrics} We use the accuracy on the testing dataset (Test Acc.) to evaluate the utility of the attacked models. We mark cases where the test accuracy drops significantly ($>10\%$) as failed attacks. Different metrics are applied to evaluate the effectiveness of different dataset auditing methods, as follows.
\begin{itemize}
    \item \textbf{Score:} \emph{Score} is the average confidence value outputted by the attack models in MIA and Rapid. A higher \emph{score} indicates that the data is more likely to be a member (trained instance) of the training dataset, and the score $>0.5$ indicates a signal of a  `trained' dataset.
    \item \textbf{Diff:} \emph{Diff} is a metric for DI. \emph{Diff} signifies the average distance between a sample and a neighbouring target class (\ie, decision boundary). A higher \emph{Diff} indicates that the model is more overfitted to the dataset, suggesting that this dataset is more likely to be a part of the training data.
    \item \textbf{Cost:} We utilize \emph{Cost (logits)} and \emph{Cost (label)} to reflect the effectiveness of using the logits or only the labels of the model for DUA. \emph{Cost (\%)} is the percentage of samples required for the audit to confidently determine success. A lower \emph{cost} indicates a more confident judgment, and a $100\%$ cost signifies a signal of `non-trained'.
    \item \textbf{WSR:} \emph{WSR} (Watermark Success Rate) measures the percentage of validation data getting the expected prediction, which is defined differently depending on the method: a specific label (DVBW/ZeroMark), an incorrect label (UBW-P/C), or the correct label (DW). We mark WSR $<25\%$ as a signal of a `non-trained' dataset.
    \item \textbf{P-value:} The \emph{p-value} is the output of the hypothesis test used in dataset auditing. The null hypothesis typically assumes that the dataset is independent of the suspicious model, while the alternative hypothesis suggests that they are related. A lower p-value indicates that the null hypothesis is less likely to hold, meaning that the dataset is more likely to be a part of the training dataset. A large number of works depend on validating whether the p-value is lower than $0.05$ to make the decision of auditing.
\end{itemize}



\begin{table*}[t]
    \centering
    \tabcolsep=1mm
    \renewcommand{\arraystretch}{1.05}
    \caption{Results of representative evasion attacks against transformer-based models trained on ImageNet. We highlight cells corresponding to failed audits in \failurecolor and those indicating successful audits in \successcolor. (Better viewed in color)}
    \label{tab:imagenet}
    \scalebox{0.70}{
    \begin{tabular}{ccccccccc} 
        \toprule 
        Type & Auditing Method & Metric & No Attack & AutoEncoder & AdvTraining-FGSM & AdvTraining-Hybrid & RS & Reprogramming\\
        \midrule 
        \multirow{10}{*}{IF-based} 
         & \multirow{2}{*}{MIA} &Test Acc. ($\%, \uparrow$) & \cellgreen 74.88 & \cellgreen 72.76 & \cellgreen 64.54 & \cellgreen 65.56 & \cellgreen 64.70 & \cellred 69.84 \\
         & & Score ($\uparrow$) & \cellgreen 0.86 & \cellgreen 0.86 & \cellgreen 0.72 & \cellgreen 0.72 & \cellred 0.00 & \cellred 0.00 \\
        \cmidrule(lr){2-9} 
         & \multirow{2}{*}{Rapid} &Test Acc. ($\%, \uparrow$) & \cellgreen 74.88 & \cellgreen 72.76 & \cellgreen 64.54 & \cellgreen 65.56 & \cellgreen 61.58 & \cellred 69.54 \\
         & & Score ($\uparrow$) & \cellgreen 0.98 & \cellgreen 0.98 & \cellgreen 0.88 & \cellgreen 0.91 & \cellred 0.00 & \cellred 0.00 \\
        \cmidrule(lr){2-9}
         & \multirow{3}{*}{DI} &Test Acc. ($\%, \uparrow$) & \cellgreen 74.88 & \cellgreen 72.76 & \cellgreen 64.54 & \cellgreen 65.56 & \cellgreen 63.20 & \cellred 68.26 \\
         & & p-value ($\downarrow$) & \cellgreen $10^{-4}$ & \cellgreen $10^{-12}$ & \cellgreen $10^{-4}$ & \cellgreen $10^{-3}$ & \cellgreen $10^{-4}$ & \cellred 0.07 \\
         & & Diff ($\uparrow$) & \cellgreen 0.62 & \cellgreen 0.61 & \cellgreen 0.50 & \cellgreen 0.69 & \cellgreen 0.50 & \cellred 0.31 \\
        \cmidrule(lr){2-9}
         & \multirow{3}{*}{DUA} &Test Acc. ($\%, \uparrow$) & \cellgreen 75.74 & \cellgreen 73.18 & \cellgreen 64.36 & \cellgreen 66.70 & \cellgreen 70.52 & \cellgreen 69.70 \\
         & & Cost (logits) ($\%, \downarrow$) & \cellgreen 30.60 & \cellgreen 69.23 & \cellred 100.00 & \cellgreen 99.80 & \cellgreen 47.14 & \cellgreen 35.63  \\
         & & Cost (label) ($\%, \downarrow$) & \cellgreen 60.46 & \cellgreen 88.23 & \cellred 100.00 & \cellred 100.00 & \cellgreen 83.63 & \cellgreen 86.20 \\
         \midrule 
        \multirow{14}{*}{EF-based}
        & \multirow{3}{*}{DVBW} &Test Acc. ($\%, \uparrow$) & \cellgreen 73.94  & \cellgreen 70.28  & \cellgreen 62.82 & \cellgreen 65.08 & \cellgreen 60.16 & \cellred 67.66 \\
         & & p-value ($\downarrow$) &\cellgreen0.00  & \cellgreen 0.00&\cellred0.89  &\cellgreen0.00 &\cellred 1.00&\cellred 1.00  \\
         & & WSR ($\%, \uparrow$) &\cellgreen100.00  &\cellgreen91.30  &\cellred37.69  &\cellgreen100.00 &\cellred 100.00& \cellred0.91  \\
        \cmidrule(lr){2-9}
         & \multirow{3}{*}{UBW-P} & Test Acc. ($\%, \uparrow$) & \cellgreen 73.62  & \cellred 71.12 & \cellgreen 62.04  & \cellgreen 64.40  & \cellgreen 62.80 & \cellred 67.80 \\
         & & p-value ($\downarrow$) &\cellgreen0.00  & \cellred0.00 &\cellgreen 0.00 &\cellgreen 0.00&\cellred1.00 &\cellred 1.00  \\
         & & WSR ($\%, \uparrow$) &\cellgreen98.34  & \cellred51.64 &\cellgreen 98.28 &\cellgreen98.44 &\cellred 97.86&\cellred 34.81 \\
        \cmidrule(lr){2-9}
         & \multirow{3}{*}{UBW-C} & Test Acc. ($\%, \uparrow$) & \cellred 73.36 & \cellred 71.54 & \cellred 64.50 & \cellred 66.38 & \cellred 64.22 & \cellred 66.34 \\
         & & p-value ($\downarrow$) &\cellred1.00  & \cellred 1.00&\cellred1.00  &\cellred1.00  &\cellred  1.00&\cellred1.00 \\
         & & WSR ($\%, \uparrow$) &\cellred20.00  &\cellred 13.92 & \cellred16.46  &\cellred 11.39 &\cellred36.70  &\cellred17.72  \\
        \cmidrule(lr){2-9}
         & \multirow{3}{*}{ZeroMark} &Test Acc. ($\%, \uparrow$) & \cellgreen 71.50 & \cellgreen 69.00 & \cellgreen 59.00 & \cellgreen 60.98 & \cellgreen 51.50 & \cellred 66.60 \\
         & & p-value ($\downarrow$) & \cellgreen 0.01 & \cellgreen 0.01 & \cellred 0.79 &  \cellgreen 0.03 & \cellgreen $10^{-5}$ & \cellred 1.00 \\
         & & WSR ($\%, \uparrow$) & \cellgreen 97.6 & \cellgreen 87.69 & \cellred 73.73 & \cellgreen 100.00 & \cellgreen 100.00 & \cellred 93.51 \\
        \cmidrule(lr){2-9}
         & \multirow{2}{*}{DW} &Test Acc. ($\%, \uparrow$) & \cellred 73.94  & \cellred 72.34 & \cellred 64.00 & \cellred 65.46  & \cellred 70.40 & \cellred 67.62 \\
         & & WSR ($\%, \uparrow$) &\cellred 0.00 & \cellred 18.00 &\cellred 0.00 &\cellred 1.99 &\cellred 0.00 &\cellred  0.00  \\
        \bottomrule 
    \end{tabular}
    }
\end{table*}

\subsection{Results of Evasion Attacks}
\label{sec:evasion}

Tables~\ref{tab:preprocessing}-\ref{tab:post} show the results of evasion attacks against ResNet-18 in the pre-processing, training, and post-training phases, respectively. Table~\ref{tab:imagenet} shows the results of representative evasion attacks against MobileViT~\citep{mehta2021mobilevit} trained on (a 100-class subset of) ImageNet~\citep{deng2009imagenet}. We maintain the realistic assumption that if an auditor needed to train auxiliary models for their auditing method (\eg, shadow models for MIA), they would use a standard architecture (ResNet-18), thus creating an architectural mismatch with the ViT suspect model. We summarize our key findings as follows.

\begin{takeawaybox}
\textbf{Takeaway 1:} Existing dataset auditing methods are generally vulnerable to evasion attacks. 
\end{takeawaybox}

The experimental results demonstrate that both IF-based and EF-based dataset auditing methods exhibit significant vulnerability when subjected to adversarial evasion attacks across different phases. Across the three tables, numerous orange cells (indicating audit failure) are distributed across various attack phases, covering all evaluated auditing methods. This indicates that the effectiveness of existing dataset auditing techniques is severely challenged when facing adversaries with evasion capabilities. For instance, in the preprocessing phase, the AutoEncoder attack and the Data Synthesis attack completely invalidate 5 and 4 different auditing methods, and WaveletFilter evades 7 of them. In the training phase, several defense mechanisms (like AdvTraining-Hybrid, ASD, DP-SGD) cause multiple IF-based and EF-based methods (\eg, DUA, DI, MIA, Rapid, UBW-C, ZeroMark, and DW under certain attacks) to fail. In the post-training phase, the Reprogramming attack also successfully evades almost all the auditing methods (except DUA).

\begin{takeawaybox}
    \textbf{Takeaway 2:} The attack effectiveness targeting a specific class of auditing methods depends on the employed strategy, but cross-class influence is also observed.
\end{takeawaybox}

The effectiveness of an attack primarily relies on the strategy. For instance, advanced decoupling attacks (\eg, Data Synthesis and DP-SGD) are effective against IF-based methods, and removal attacks (\eg, WaveletFilter and Adversarial Training) work against EF-based methods. In the meantime, cross-class impacts of attacks also exist. WaveletFilter attacks successfully evade DI and MIA. The Reprogramming attack even has a significant negative impact on 8 of 9 methods (including both auditing types). These results demonstrate that attacks primarily designed against one class of methods can also exhibit unintended effectiveness against the other.


\begin{takeawaybox}
    \textbf{Takeaway 3:} EF-based methods relying on strong external features show relatively higher robustness but are not immune to evasion attacks.
\end{takeawaybox}

\begin{table*}[t]
    \centering
    \tabcolsep=1mm
    \renewcommand{\arraystretch}{1.05}
    \caption{Results of representative evasion attacks in the fine-tuning scenario. We highlight cells corresponding to failed audits in \failurecolor{} and those indicating successful audits in \successcolor. (Better viewed in color)}
    \label{tab:finetuning}
    \scalebox{0.70}{
    \begin{tabular}{ccccccccc} 
        \toprule 
        Type & Auditing Method & Metric & No Attack & AutoEncoder & AdvTraining-Hybrid & DP-SGD ($\epsilon=64$) & RS & Reprogramming\\
        \midrule 
        \multirow{10}{*}{IF-based} 
        & \multirow{2}{*}{MIA} &Test Acc. ($\%, \uparrow$) & \cellgreen 88.26 & \cellgreen 85.70 & \cellgreen 85.60 & \cellred 72.87 & \cellred 85.70 & \cellred 84.96 \\
         & & Score ($\uparrow$) & \cellgreen 0.62 & \cellgreen 0.59 & \cellgreen 0.56 & \cellred 0.52 & \cellred 0.00 & \cellred 0.00 \\
        \cmidrule(lr){2-9} 
         & \multirow{2}{*}{Rapid} &Test Acc. ($\%, \uparrow$) & \cellgreen 88.26 & \cellgreen 85.70 & \cellgreen 85.60 & \cellred 72.87 & \cellred 85.70 & \cellred 86.24 \\
         & & Score ($\uparrow$) & \cellgreen 0.62 & \cellgreen 0.65 & \cellgreen 0.68 & \cellred 0.48 & \cellred 0.33 & \cellred 0.27 \\
        \cmidrule(lr){2-9}
         & \multirow{3}{*}{DI} &Test Acc. ($\%, \uparrow$) & \cellred 88.26 & \cellred 85.70 & \cellred 85.60 & \cellgreen 72.87 & \cellgreen 85.70 & \cellred 85.71 \\
         & & p-value ($\downarrow$) & \cellred 0.06 & \cellred 0.55 & \cellred 0.10 & \cellgreen 0.05 & \cellgreen $10^{-4}$ & \cellred 0.11 \\
         & & Diff ($\uparrow$) & \cellred 0.27 & \cellred 0.00 & \cellred 0.00 & \cellgreen 0.10 & \cellgreen 1.14 & \cellred 0.00 \\
        \cmidrule(lr){2-9}
         & \multirow{3}{*}{DUA} &Test Acc. ($\%, \uparrow$) & \cellgreen 87.87 & \cellgreen 85.44 & \cellgreen 85.33 & \cellgreen 72.25 & \cellgreen 86.90 & \cellred 77.61 \\
         & & Cost (logits) ($\%, \downarrow$) & \cellgreen 23.77 & \cellgreen 21.97 & \cellgreen 89.97 & \cellgreen 99.06 & \cellgreen 34.46 & \cellred 100.00  \\
         & & Cost (label) ($\%, \downarrow$) & \cellgreen 79.03 & \cellgreen 84.60 & \cellgreen 99.91 & \cellgreen 89.09 & \cellgreen 89.71 & \cellred 100.00 \\
         \midrule 
        \multirow{14}{*}{EF-based}
        & \multirow{3}{*}{DVBW} &Test Acc. ($\%, \uparrow$) & \cellgreen 87.50 & \cellred 83.83 & \cellgreen 83.89 & \cellgreen 67.84 & \cellred 63.45 & \cellred 75.90 \\
         & & p-value ($\downarrow$) &\cellgreen 0.00 & \cellred1.00 &\cellgreen 0.00 &\cellgreen0.00 &\cellred1.00 &\cellred 1.00  \\
         & & WSR ($\%, \uparrow$) &\cellgreen 100.00 &\cellred 20.01 &\cellgreen 99.92&\cellgreen100.00 &\cellred 100.00& \cellred 10.74 \\
        \cmidrule(lr){2-9}
         & \multirow{3}{*}{UBW-P} & Test Acc. ($\%, \uparrow$) & \cellgreen 87.66 & \cellred 84.39 & \cellgreen 83.10 & \cellgreen 68.07 & \cellred 80.06 & \cellred 85.71 \\
         & & p-value ($\downarrow$) &\cellgreen0.00  & \cellred1.00 &\cellgreen 0.00 &\cellgreen0.00&\cellred1.00 &\cellred 1.00  \\
         & & WSR ($\%, \uparrow$) &\cellgreen86.93  & \cellred 17.37&\cellgreen  82.92&\cellgreen84.69 &\cellred87.80 &\cellred15.20  \\
        \cmidrule(lr){2-9}
         & \multirow{3}{*}{UBW-C} & Test Acc. ($\%, \uparrow$) & \cellgreen 86.74 & \cellgreen 84.90 & \cellgreen 84.68 & \cellred 68.75 & \cellred 78.70 & \cellred 84.56 \\
         & & p-value ($\downarrow$) &\cellgreen $10^{-143}$ & \cellgreen $10^{-148}$ &\cellgreen $10^{-90}$ &\cellred1.00 &\cellred1.00  &\cellred 1.00\\
         & & WSR ($\%, \uparrow$) &\cellgreen 76.24 &\cellgreen 77.98 & \cellgreen  65.04&\cellred 25.49 &\cellred 98.63 &\cellred76.99  \\
        \cmidrule(lr){2-9}
         & \multirow{3}{*}{ZeroMark} &Test Acc. ($\%, \uparrow$) & \cellgreen 87.81 & \cellred 83.39 & \cellred 83.19 & \cellred 71.45 & \cellred 65.26 & \cellred 85.56 \\
         & & p-value ($\downarrow$) & \cellgreen $10^{-98}$ & \cellred 1.00 & \cellred 1.00 &  \cellred 1.00 & \cellred 1.00 & \cellred 1.00 \\
         & & WSR ($\%, \uparrow$) & \cellgreen 100.00 & \cellred 15.72 & \cellred 99.64 & \cellred 100.00 & \cellred 100.00 & \cellred 10.68 \\
        \cmidrule(lr){2-9}
         & \multirow{2}{*}{DW} &Test Acc. ($\%, \uparrow$) & \cellgreen 87.74 & \cellred 85.52 & \cellred 84.83 & \cellred 67.90 & \cellgreen 86.99 & \cellgreen 73.49 \\
         & & WSR ($\%, \uparrow$) &\cellgreen 68.60 & \cellred 16.50 &\cellred0.50  &\cellred 0.50 &\cellgreen 69.70 &\cellgreen  70.00  \\
        \bottomrule 
    \end{tabular}
    }
    \vspace{-0.4em}
\end{table*}

A key finding is that dataset auditing methods relying on the embedding of strong external features, exemplified by DVBW and UBW-P in our experiments, exhibit relatively higher robustness. Their robustness is particularly exhibited against challenging training-phase attacks like adversarial training and DP-SGD. The two methods are poison-label methods that necessitate changing the labels of part of the samples in the dataset. However, even these methods are not immune to evasion attacks, showing vulnerabilities to specific pre-processing data transformations (\eg, WaveletFilter, AutoEncoder) and post-training inference manipulations (\eg, Reprogramming). 
Besides, other EF-based methods tested using mild external features (\ie, UBW-C, ZeroMark, DW) displayed considerably less robustness across various attack stages.

Moreover, IF-based methods generally exhibit moderate robustness, with DUA demonstrating relatively higher robustness. DUA enhances auditing effectiveness by injecting additional noise into the dataset. This mechanism improves robustness but also makes the audit results more susceptible to forgery attacks. We will further discuss the details in Section~\ref{sec:expforgery}.

\begin{takeawaybox}
    \textbf{Takeaway 4}: For modern models and large-scale datasets, the vulnerability of dataset auditing still holds.
\end{takeawaybox}
Consistent with our primary findings, the results in Table~\ref{tab:imagenet} demonstrate that existing dataset auditing methods remain highly vulnerable to evasion attacks when auditing modern models (Vision Transformers) and large-scale datasets (ImageNet). Both IF-based and EF-based auditing techniques frequently failed to correctly identify the usage of the ImageNet victim dataset when attacks like AutoEncoder-based denoising, adversarial training, or Reprogramming were employed. The architectural mismatch (transformer-based model and auditor's CNN) and the increased dataset complexity do not confer robustness to the evaluated auditing methods.

\begin{table*}[t]
    \centering
    \tabcolsep=4mm
    \renewcommand{\arraystretch}{1.05}
    \caption{Results of forgery attacks in the white-box scenario. We highlight cells corresponding to failed audits in \failurecolor and those indicating successful audits in \successcolor. (Better viewed in color)}
    \label{tab:forgery}
    \scalebox{0.70}{
    \begin{tabular}{ccccccccc} 
        \toprule 
        Type & Auditing Method & Metric & No Attack & FGSM & PGD & UAP & TIFGSM & VNIFGSM\\
        \midrule 
        \multirow{6}{*}{IF-based}
        & \multirow{1}{*}{MIA} & Score ($\downarrow$) & \cellgreen 0.54 & \cellred 0.66 & \cellred 0.72 & \cellgreen 0.54 & \cellred 0.71 & \cellred 0.71 \\
        \cmidrule(lr){2-9} 
        & \multirow{1}{*}{Rapid} & Score ($\downarrow$) & \cellgreen 0.22 & \cellred 0.79 & \cellred 0.89 & \cellgreen 0.30 & \cellred 0.86 & \cellred 0.88  \\
        \cmidrule(lr){2-9}
        & \multirow{2}{*}{DI} & p-value ($\uparrow$) & \cellred 0.01 & \cellred $10^{-13}$ & \cellred $10^{-5}$ & \cellred $10^{-23}$ & \cellgreen 0.97 & \cellred 0.01 \\
        &  & Diff ($\downarrow$) & \cellred 0.79 & \cellred 1.35 & \cellred 0.98 & \cellred 1.61 & \cellgreen -0.81 & \cellred 0.09 \\
        \cmidrule(lr){2-9}
        & \multirow{2}{*}{DUA} & Cost (logits) (\%, $\uparrow$) & \cellgreen 100.00 & \cellred 0.13 & \cellred 0.13 & \cellred 0.18 & \cellred 0.13 & \cellred 0.24 \\
        &  & Cost (label) (\%, $\uparrow$) & \cellgreen 100.00 & \cellred 0.13 & \cellred 0.13 & \cellred 0.24 & \cellred 0.13 & \cellred 0.24 \\
         \midrule 
         \multirow{7}{*}{EF-based}
        & \multirow{2}{*}{DVBW} & p-value ($\uparrow$) & \cellgreen 1.00 & \cellred 0.00 & \cellred 0.00 &\cellred 0.00 & \cellred 0.00 & \cellred 0.00 \\
        &  & WSR (\%, $\downarrow$) & \cellgreen 10.01 & \cellred 29.92 & \cellred 98.37 &\cellred 65.60 & \cellred 69.77 &\cellred 96.77 \\
        \cmidrule(lr){2-9}
        & \multirow{2}{*}{UBW} & p-value ($\uparrow$) &\cellgreen 1.00 &\cellred 0.00  &\cellred 0.00 &\cellred 0.00 & \cellred 0.00 & \cellred 0.00 \\
        &  & WSR (\%, $\downarrow$) &\cellgreen 11.73 &\cellred 86.55  & \cellred 100.00 &\cellred 71.08 & \cellred 95.69 &\cellred 100.00\\
        \cmidrule(lr){2-9}
        & \multirow{2}{*}{ZeroMark} & p-value ($\uparrow$) & \cellgreen 1.00 & \cellgreen 1.00 & \cellgreen 1.00 & \cellgreen 0.99 & \cellred 0.01 & \cellgreen 1.00 \\
        &  & WSR (\%, $\downarrow$) & \cellgreen 10.01 & \cellgreen 29.93 & \cellgreen 98.44 & \cellgreen 68.28 & \cellred 71.68 & \cellgreen 96.88 \\
        \cmidrule(lr){2-9}
        & \multirow{1}{*}{DW} & WSR (\%, $\downarrow$) & \cellgreen 11.73 & \cellred 67.00 & \cellred 89.40 & \cellred 66.37 &\cellred 73.72  &\cellred 89.36 \\
        \bottomrule 
    \end{tabular}
    }
    \vspace{-0.5em}
\end{table*}

\begin{table*}[t]
    \centering
    \tabcolsep=4mm
    \renewcommand{\arraystretch}{1.05}
    \caption{Results of forgery attacks in the black-box scenario. We highlight cells corresponding to failed audits in \failurecolor and those indicating successful audits in \successcolor. (Better viewed in color)}
    \label{tab:forgery-blackbox}
    \scalebox{0.70}{
    \begin{tabular}{ccccccccc} 
        \toprule 
        Type & Auditing Method & Metric & No Attack & FGSM & PGD & UAP & TIFGSM & VNIFGSM\\
        \midrule 
        \multirow{6}{*}{IF-based}
        & \multirow{1}{*}{MIA} & Score ($\downarrow$) & \cellgreen 0.54 & \cellgreen 0.54 & \cellred 0.55 & \cellgreen 0.53 & \cellred 0.55 & \cellred 0.55 \\
        \cmidrule(lr){2-9} 
        & \multirow{1}{*}{Rapid} & Score ($\downarrow$) & \cellgreen 0.22 & \cellgreen 0.16 & \cellgreen 0.28 & \cellgreen 0.15 & \cellgreen 0.25 & \cellgreen 0.13  \\
        \cmidrule(lr){2-9}
        & \multirow{2}{*}{DI} & p-value ($\uparrow$) & \cellred 0.01 & \cellred $10^{-48}$ & \cellred $10^{-72}$ & \cellred $10^{-14}$ & \cellgreen 0.88 & \cellred $10^{-4}$ \\
        &  & Diff ($\downarrow$) & \cellred 0.79 & \cellred 1.86 & \cellred 1.97 & \cellred 1.53 & \cellgreen -0.42 & \cellred 0.41 \\
        \cmidrule(lr){2-9}
        & \multirow{2}{*}{DUA} & Cost (logits) (\%, $\uparrow$) & \cellgreen 100.00 & \cellred 0.13 & \cellred 0.37 & \cellred 0.41 & \cellred 0.19 & \cellred 0.13 \\
        &  & Cost (label) (\%, $\uparrow$) & \cellgreen 100.00 & \cellred 1.27 & \cellred 1.71 & \cellred 0.69 & \cellred 0.19 & \cellred 0.13 \\
        \midrule 
        \multirow{7}{*}{EF-based}
        & \multirow{2}{*}{DVBW} & p-value ($\uparrow$) & \cellgreen 1.00 & \cellred 0.00 & \cellred 0.00 &\cellred 0.00 & \cellred $10^{-144}$ & \cellred 0.00 \\
        &  & WSR (\%, $\downarrow$) & \cellgreen 10.01 & \cellred 16.87 & \cellred 18.53 &\cellred 35.12 & \cellred 23.48 &\cellred 33.25 \\
        \cmidrule(lr){2-9}
        & \multirow{2}{*}{UBW} & p-value ($\uparrow$) &\cellgreen 1.00 &\cellred 0.00  &\cellred 0.00 &\cellred 0.00 & \cellred $10^{-236}$ & \cellred 0.00 \\
        &  & WSR (\%, $\downarrow$) &\cellgreen 11.73 &\cellred 21.73  & \cellred 22.29 &\cellred 35.99 & \cellred 29.44 &\cellred 37.95\\
        \cmidrule(lr){2-9}
        & \multirow{2}{*}{ZeroMark} & p-value ($\uparrow$) & \cellgreen 1.00 & \cellgreen 0.99 & \cellgreen 1.00 & \cellgreen 0.99 & \cellgreen 1.00 & \cellgreen 1.00 \\
        &  & WSR (\%, $\downarrow$) & \cellgreen 10.01 & \cellgreen 10.36 & \cellgreen 10.73 & \cellgreen 10.81 & \cellgreen 11.49 & \cellgreen 12.02 \\
        \cmidrule(lr){2-9}
        & \multirow{1}{*}{DW} & WSR (\%, $\downarrow$) & \cellgreen 11.73 & \cellgreen 20.68 & \cellgreen 19.49 & \cellred 35.15 &\cellgreen 25.28  &\cellred 34.35 \\
        \bottomrule 
    \end{tabular}
    }
\end{table*}

\subsection{Evaluating Dataset Auditing in the Fine-tuning Scenario}
\label{apd:finetuning}


Beyond training models from scratch, fine-tuning pre-trained models on a task-specific dataset is a common and efficient paradigm in modern DL. We simulate the fine-tuning scenario using a ResNet-18 pre-trained on ImageNet-1000 and then fine-tune it using CIFAR-10 as the target (victim) dataset. Other experimental settings remain consistent with our main experiments in Section~\ref{sec:evasion}.

The experimental results of applying evasion attacks in the fine-tuning scenario are presented in Table~\ref{tab:finetuning}. The results indicate that dataset auditing methods exhibit vulnerabilities similar to those observed when training from scratch. Both IF-based methods and EF-based methods frequently fail when confronted with evasion attacks such as DP-SGD, RS, or Reprogramming. 
This suggests that the inherent weaknesses of existing dataset auditing techniques are not limited to models trained from scratch but also extend to the fine-tuning paradigm.

\subsection{Results of Forgery Attacks}
\label{sec:expforgery}

For forgery attacks, we assume that the adversary wants to make their own dataset (\ie, the attack auxiliary dataset) to be falsely identified as the `trained dataset' of other independent models. We use the model trained on the unmarked release dataset as the independent model. The results of white-box forgery attacks are in Table~\ref{tab:forgery}, and the experiments in the black-box scenario can be found in Table~\ref{tab:forgery-blackbox}.

\begin{figure*}[t]
    \centering
    \includegraphics[width=0.98\linewidth]{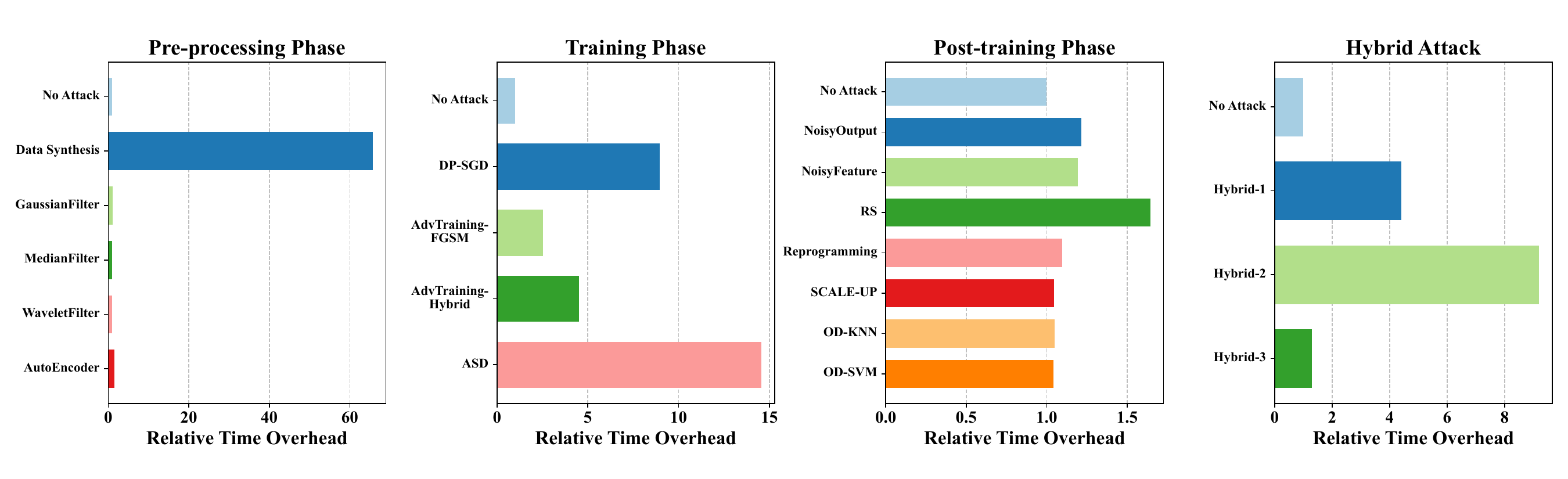}
    \caption{Relative time overhead of different evasion attacks. The overhead is compared with the baseline time without any specific evasion attack (\ie, `No Attack' with a relative overhead normalized to 1.0).}
    \label{fig:overhead}
\end{figure*}

\begin{takeawaybox}
    \textbf{Takeaway 5:} Existing dataset auditing methods are overall vulnerable to forgery attacks.
\end{takeawaybox}

Our experimental results strongly support this finding. When subjected to various forgery attacks, nearly all evaluated auditing methods exhibited significant failures. For instance, IF-based methods, which rely on detecting subtle differences in model behavior (like prediction confidence or loss), were readily deceived. The forgery attacks successfully manipulated model outputs on the forged samples to resemble those of genuinely trained data points, causing these methods to incorrectly flag the unused dataset as `trained' (indicated by drastically reduced costs for DUA, low p-values, and large Diff for DI, and increased scores for MIA/Rapid). Similarly, EF-based methods, including DVBW, UBW, and DW, which look for specific EFs, are also compromised. The forgery attacks effectively generated inputs that triggered the expected model behaviors, leading to low p-values and high Watermark Success Rates (WSR), thus falsely indicating the presence of the EFs and confirming usage. 

\begin{takeawaybox}
    \textbf{Takeaway 6:} Dataset auditing methods that involve more complex decision-making and auditing processes are more resistant to forgery.
\end{takeawaybox}

While most methods proved vulnerable, those employing more intricate verification mechanisms demonstrated comparatively better resilience. For example, different from other EF-based methods that directly calculate the accuracy of some specific trigger samples, ZeroMark analyzes the boundary gradients of the closest boundary samples. This sophisticated auditing process makes the auditing results hard to forge. Therefore, although the forgery attack improves the WSR, it only succeeds once against ZeroMark, with the p-values in all other cases remaining close to $1$. Methods relying on simpler signals, like raw confidence scores or basic misclassification, are demonstrably easier to deceive.





\subsection{Attack Overhead}

Beyond evaluating the effectiveness of evasion attacks, understanding their computational overhead is crucial for assessing their practical feasibility for an adversary. An adversary might be constrained by computational resources, potentially limiting their choice of attacks. To quantify this, we measured the execution time required for applying each evasion attack implemented in \sys during the model training or deployment process, using our primary experimental setup. For each attack, we evaluate the time overhead of the entire pipeline, including data preprocessing, training, and post-training model testing. Figure~\ref{fig:overhead} visualizes the relative time overhead of each attack compared to the baseline time without any specific evasion attack (represented as `No Attack' with a relative overhead normalized to 1.0). The overheads are evaluated using one NVIDIA RTX 3090 GPU.

Our analysis, depicted across the four panels of Figure 3 corresponding to the different attack phases, reveals significant variations in computational cost. 
\begin{itemize}
    \item In the pre-processing phase, most of the attacks (except data synthesis) exhibit very similar overhead. Data synthesis stands out as exceptionally expensive. It incurs an overhead approximately 65 times greater than baseline. This substantial cost is primarily due to the necessity of training a diffusion model and generating synthesized data using the trained generative model.
    \item  In the training phase, integrating evasion mechanisms directly into the training loop generally introduces noticeable overhead. Attacks during the training phase incur at least a 1.5× and at most a 13.6× increase in time overhead compared to the baseline. 
    \item In the post-training phase, all the attacks are consistently lightweight. Most methods demonstrate relative time overheads close to $1.0$. This is expected, as these techniques typically involve minor additions or modifications to the standard inference pipeline, which is computationally much less demanding than model training.
\end{itemize}

\begin{figure}[t]
    \centering
    \includegraphics[width=0.99\linewidth]{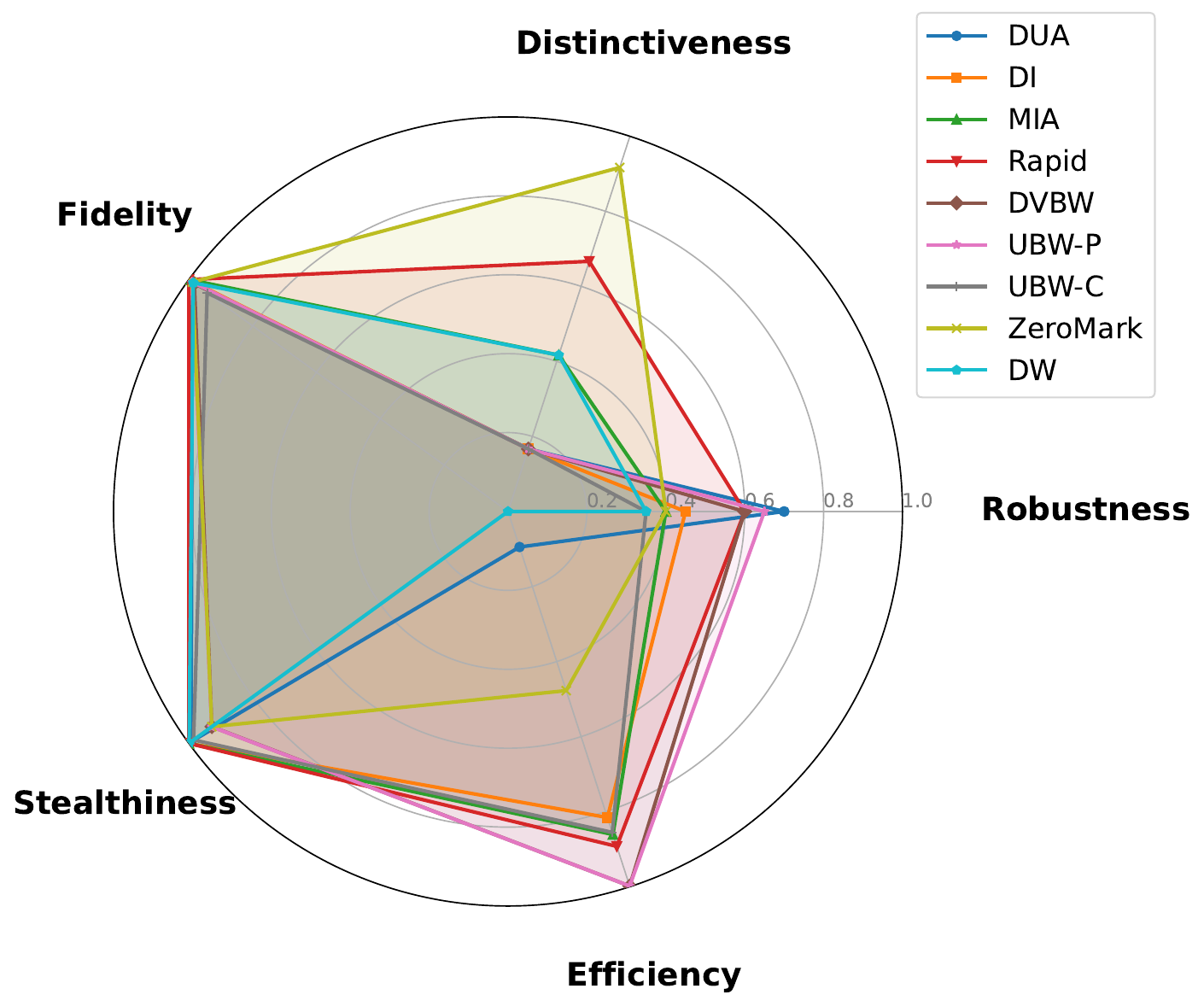}
    \caption{The summary of evaluated dataset auditing methods. This figure overall illustrates the relative performance of the 9 evaluated dataset auditing methods across 5 dimensions.}
    \label{fig:compare}
\end{figure}

\subsection{Summary of Evaluated Dataset Auditing Methods}

To provide a holistic comparison of the evaluated dataset auditing methods, we summarize their relative performance across five key dimensions: Robustness, Distinctiveness, Fidelity, Stealthiness, and Efficiency according to the design objectives in Section~\ref{sec:objective}. Figure~\ref{fig:compare} represents this comparison using a radar chart, where scores closer to the outer edge (1.0) indicate better performance in that dimension. 

Our analysis reveals that no single method excels across all 5 dimensions. The most critical finding is the pervasive lack of adversarial robustness and distinctiveness, as also analyzed in the previous section. On the other hand, fidelity is typically high, indicating minimal impact on model utility in benign settings. A clear divide exists in stealthiness. IF-based methods inherently excel in stealthiness, while EF-based methods depend on the embedded feature. Efficiency spans a wide spectrum, from computationally cheap methods to highly demanding ones requiring intensive processing overhead. 

In conclusion, the comparative analysis highlights a critical gap: current dataset auditing techniques force difficult trade-offs between essential properties. The widespread vulnerability to evasion attacks and forgery attacks, in particular, highlights a critical future direction.

\section{Discussion}
\label{sec:discussion}

\subsection{Related Works}

Since dataset auditing in DL is an emerging topic, there are only a limited number of surveys or benchmarks that have explicitly explored this domain. Asswad et al.\citep{asswad2021data} highlighted the importance of protecting data ownership and reviewed representative related techniques. Chandrasekaran et al.\citep{chandrasekaran2021sok} provided a comprehensive study on machine learning governance, including data ownership in machine learning. Recently, Du et al.~\citep{du2025sok} presented the first systematic SoK regarding dataset auditing, primarily focusing on surveying existing solutions. From the broader perspective of AI copyright protection, the majority of existing work focuses on watermarking techniques for models~\citep{ren2024copyright, shao2025sok}, generated content~\citep{ren2024sok, zhao2024sok}, and traditional multimedia~\citep{liu2024survey}. Lukas et al.~\citep{lukas2022sok} presented a toolbox to evaluate the robustness of DNN watermarking, which is fundamentally different from dataset auditing.

Compared with existing work, our paper makes the following four distinct contributions to the field. \textbf{(1)} This paper focuses on a specific scenario (\ie, auditing the usage of a dataset) where existing techniques to watermark other objects are not feasible. The dataset is specifically used for training a DL model, and the model is the audit target. \textbf{(2)} This paper aims to provide a novel adversarial angle for evaluating existing dataset auditing methods. \textbf{(3)} Existing works fail to provide a comprehensive benchmark and evaluation regarding the SOTA dataset auditing methods. \textbf{(4)} This paper investigates new attack vectors, such as forgery attacks against dataset auditing, which are rarely discussed in prior papers~\citep{zhu2025evading}.


\subsection{Dataset Auditing for Generative Models}
\label{sec:generative}
While dataset auditing has been actively explored for classification models, auditing generative models has only recently started to receive attention \citep{qiu2025watermarking,hu2025membership,li2025towards} and remains largely underdeveloped. Many recent results suggest that even achieving stable results useful for auditing in real-world non-adversarial settings is extremely challenging \citep{duan2023diffusion,duanmembership}. This difficulty arises from several inherent characteristics of generative models: \textbf{(1)} They are typically trained on massive-scale datasets (often containing billions of samples), where any individual data subset constitutes only a minuscule fraction, making both internal and external auditing signals significantly diluted. \textbf{(2)} The generation process is highly stochastic and sensitive to sampling seeds, further complicating reliable attribution. \textbf{(3)} Many real-world misuse cases manifest at the output level (\eg, a generated image resembling a copyrighted artwork), where neither internal model parameters nor standard interfaces are available (\ie, the box-free setting). In such cases, traditional auditing methods, which rely on model access or specific behavioral signatures, become ineffective or even inapplicable. While very recent works have achieved partial success in specific settings (\eg, in fine-tuning scenarios \citep{qiu2025watermarking,he2025towards,li2025towards}), designing stable and general-purpose auditing algorithms for generative models remains an open challenge in the community.

Given these complexities, existing auditing methods for generative models are not yet mature and may yield unstable or misleading results when evaluated under adversarial settings. Therefore, we caution that rushing into adversarial robustness evaluations in such an ill-defined setting without reliable baselines may misrepresent the true difficulty of the problem or misguide the community toward premature conclusions. Nevertheless, our framework and attack methodology—structured by three processing stages (\ie, pre-processing, training, post-training) and three attack strategies (\ie, decoupling, removal, detection)—are designed to be model-agnostic and generalize across different modalities. We leave the extension to generative models as an important direction for future work.

\subsection{Potential Future Direction}

\firstpartitle{Future Direction for Dataset Auditing} The demonstrated vulnerability of existing dataset auditing methods underscores an urgent need for fundamentally more robust and distinctive techniques. Future work may focus on developing novel auditing mechanisms resilient to these attacks. \textbf{(1)} For evasion, this could involve exploring IFs that are harder to decouple from the model utility. For EF-based methods, future works may seek EFs that are deeply integrated into the model's core functionality rather than easily removable or detectable trigger patterns. A certified robust mechanism is also urgently needed. \textbf{(2)} For forgery, our empirical study suggests that complicating the auditing process may lead to better resistance. Consequently, exploring more complex methods, such as multi-bit watermarks~\citep{shao2025explanation} or cryptography~\citep{fiege1987zero}, may be a promising future direction to improve dataset auditing.

\partitle{Limitations and Future Directions for Evaluating Dataset Auditing} While \sys provides a comprehensive benchmark, there are still some limitations. First, this paper primarily focuses on dataset auditing regarding image datasets since they are currently the most widely studied and feasible methods. There have been some preliminary works studying auditing datasets of other modalities, such as text~\citep{huang2024general, wei2024proving}, point clouds~\citep{wei2024pointncbw}, and audio~\citep{li2025cbw}. Providing a comprehensive evaluation of these works may be an important future work. Second, this paper does not cover all possible scenarios of dataset usage in DL. We primarily evaluate the two most important and widely adopted settings: training from scratch (Section~\ref{sec:evaluation}) and fine-tuning (Appendix~\ref{apd:finetuning}). Third, \sys does not thoroughly explore finer-grained threat models, such as adversaries with varying levels of capabilities (an adversary with very limited computational resources or data). An exhaustive study of the attacker's capabilities would be a meaningful direction for future work.



\section{Conclusion}
\label{sec:conclusion}

This paper presented a comprehensive adversarial evaluation of dataset auditing in DL systems, addressing the critical gap between existing auditing methods and their resilience against dedicated attacks. We introduced a novel internal feature and external feature taxonomy to classify auditing techniques, formally defined evasion and forgery attack objectives, and proposed systematic attack strategies to realize them. These contributions led to the development of \sys, a benchmark featuring 17 evasion and 5 forgery attacks. \sys also provides an extensible toolbox to design new dataset auditing methods and attacks. This toolbox can also facilitate the unified evaluation of auditing methods. Our empirical experiments of 9 representative auditing methods against \sys revealed some critical findings, \eg, none of them demonstrated sufficient robustness or distinctiveness under adversarial settings. These results highlight the significant vulnerability of current methods and underscore the urgent need for developing novel and adversarially resilient dataset auditing methods.

\bibliography{ref}

\appendix


\section{Details of Dataset Auditing in \sys}
\label{apd:detailaudit}

\tightpartitle{Shokri et al. (MIA)~\citep{shokri2017membership}}
Shokri et al. proposed the first membership inference attack (MIA), leveraging models' differential behavior (\eg, higher confidence) on members versus non-members. The method trains multiple `shadow models' to generate labeled prediction vectors (member/non-member), which then train a final classifier to distinguish target model members. We used 10 shadow models in our experiments.

\tightpartitle{Rapid~\citep{he2024difficulty}}
Rapid is a SOTA MIA addressing inherent errors in difficulty calibration methods of reference-based attacks. It combines the original membership signal (\eg, loss) with the calibrated score from reference models. Rapid trains a scoring model mapping both scores to a final membership prediction, learning to correct limitations of relying solely on calibration. We used 4 reference models in our experiments.

\tightpartitle{Dataset Inference (DI)~\citep{maini2020dataset}}
DI assesses if a suspicious model shows higher prediction margins for the owner's data points. It generates feature embeddings capturing this margin from the suspicious model using white-box (MinGD) or black-box (Blind Walk) methods. A confidence regressor, pre-trained on the owner's model and data, scores these embeddings. Finally, a statistical hypothesis test aggregates scores to determine if the suspicious model used the owner's dataset.

\tightpartitle{Data-use Auditing (DUA)~\citep{huang2024general}} 
DUA is a general dataset auditing framework. The owner creates two marked versions of each datum (preserving utility, maximizing difference), publishing one and hiding the other. Then, it uses a black-box Membership Inference (MI) score function and compares scores on the published versus hidden versions (contrastive MI). A sequential hypothesis test then determines if the published data was used for training.

\tightpartitle{DVBW~\citep{li2023black}} DVBW proposes the first and a general framework for backdoor-based dataset auditing. Specifically, DVBW utilizes poison-label backdoor attacks for dataset auditing and presents a hypothesis-test-based verification method. In our experiments, the target label is set to 1.

\tightpartitle{UBW-P/C~\citep{li2022untargeted}} UBW is proposed for harmless and stealthy dataset copyright protection using untargeted backdoors. UBW has two variants, UBW-P and UBW-C. UBW-P utilizes an untargeted poison-label backdoor attack that changes the labels of the poisoned samples randomly. In contrast, UBW-C adopts bi-level optimization to craft clean-label (\ie, the labels of the watermarked samples do not change) samples. Finally, UBW also utilizes a hypothesis test to complete dataset auditing.

\tightpartitle{ZeroMark~\citep{guo2024zeromark}} ZeroMark is proposed for secure dataset auditing without disclosing watermarks. Inspired by the intrinsic property that boundary gradients of watermarked DNNs align with the watermark pattern, ZeroMark generates boundary versions of benign samples and calculates their boundary gradients under a label-only black-box setting. It then performs a hypothesis test based on the cosine similarity between the gradients and the secret watermark pattern to achieve auditing.

\tightpartitle{Domain Watermark (DW)~\citep{guo2023domain}} DW is proposed for harmless dataset auditing by identifying a hardly-generalized domain. Instead of causing misclassification, DW ensures watermarked models correctly classify specific `hard' samples misclassified by benign models. DW employs bi-level optimization to generate the domain and craft visually indistinguishable clean-label modified samples for stealth.

\section{Details of Evasion Attacks in \sys}
\label{apd:detailevasion}

\tightpartitle{Data Augmentation} Data augmentation is a widely-used approach to apply transformations to data to increase its diversity. In our experiments, we apply random cropping, resizing, and flipping as the augmentation methods.

\tightpartitle{Data Synthesis} Data synthesis involves generating artificial data, typically using generative models trained on the original dataset, to train the suspect model, thereby obscuring the direct link to the protected data. In our experiments, we use the DDIM model~\citep{ho2020denoising} as the generative model.

\tightpartitle{Denoising Filters} Denoising filters apply traditional image processing filters (\eg, Gaussian, Median, Wavelet) to potentially remove noise-like external features embedded for auditing, effectively cleaning the data before model training. In our experiments, we set the kernel size to be 3.

\tightpartitle{Autoencoder-based Denoising} Autoencoder-based Denoising employs a trained autoencoder model to reconstruct the data and remove EFs. In our experiments, we adopt a 6-layer CNN-based autoencoder model.

\tightpartitle{Regularization} Regularization refers to the weight decay during training to prevent the model from overfitting. In our experiments, we use the classic $\ell_2$ regularization.

\tightpartitle{DP-SGD} DP-SGD adds calibrated noise during training to obscure the model's dependence on specific training examples. In our experiments, we apply the Gaussian mechanism and test two different privacy budgets, 32 and 64. 

\tightpartitle{Adversarial Training} Adversarial Training augments the training data with adversarially perturbed examples. We use two different implementations of adversarial training. AdvTraining-FGSM uses solely the FGSM method~\citep{goodfellow2014explaining} to generate adversarial examples while AdvTraining-Hybrid adopts three methods, FGSM~\citep{goodfellow2014explaining}, PGD~\citep{madry2018towards}, and C\&W~\citep{carlini2017towards}.

\tightpartitle{Adaptive Splitting Poisoned Dataset (ASD)~\citep{gao2023backdoor}}
ASD is a training-time backdoor defense. ASD employs loss-guided splitting and meta-learning-inspired splitting to dynamically update two separate data pools. By leveraging the clean and poisoned data pools obtained through this adaptive partitioning, ASD effectively mitigates backdoors during training.

\tightpartitle{Noisy Output \& Noisy Feature} Noise is injected into the model's intermediate feature activations or final output predictions during inference to mask the subtle signals. In our experiments, we use Gaussian noise and the standard deviation is set to 0.5.


\tightpartitle{Randomized Smoothing (RS)} RS smooths the model's predictions by averaging outputs over multiple noisy input perturbations. We set the number of perturbed samples to 30.

\tightpartitle{Reprogramming~\citep{chen2025refine}} The reprogramming technique modifies the model's inference process, often using learnable input and output transformations, to neutralize the specific abnormal behaviors. In our experiments, we use a 10-layer U-Net as the input transformation module and train the U-Net for 10 epochs.

\tightpartitle{SCALE-UP~\citep{guo2023scale}}
SCALE-UP is an effective black-box, input-level backdoor detection method. SCALE-UP identifies and filters malicious test samples by analyzing their prediction consistency during the pixel-wise amplification process.

\tightpartitle{Outlier Detection (OD)} OD identifies input samples that are statistically different from typical data (potentially EF triggers), allowing the adversary to return a non-standard or random output to evade detection. In \sys, we use two different models, \ie, KNN and SVM.

To ensure simplicity and fairness in our evaluation, we adopt the most classical configurations for each attack. While this may introduce some degradation in model utility in certain scenarios, we note that such impacts can often be mitigated via parameter tuning, albeit potentially at the expense of attack efficacy. Users may further adjust these parameters based on their specific needs or evaluation preferences, which, however, falls beyond the scope of this work.

\section{Details of Forgery Attacks in \sys}
\label{apd:detailforgery}

In forgery attacks, we implement different targeted adversarial attacks to craft fake `trained' samples. Specifically, for DUA, DVBW, ZeroMark, and DW. We set the target label to 4. For UBW-P/C, since UBW tests whether the testing samples are misclassified (without specifying the target label), we implement the forgery attacks using untargeted methods. For DI, MIA, and Rapid, our objective is to increase the output confidence of these forged samples. Therefore, we apply a reverse untargeted attack (\ie, in contrast to the optimization direction of untargeted attacks) to maximize confidence.

\section{Other Detailed Experimental Settings}
\label{apd:detail}

As default settings, we train ResNet-18 on CIFAR-10 from scratch for in total 90 epochs with a learning rate of 0.03. The batch size is set to 128. Moreover, since batch normalization in ResNet-18 does not support DP-SGD, we alter these layers with modern group normalization. For EF-based methods, we use the Blended method~\citep{chen2017targeted}, which applies a weighed blend of a specific trigger pattern and the original image, with the trigger pattern assigned a weight of 0.2. 

\end{document}